\documentclass{aa}  
\def\tauAmin{0.05}
\def\tauAmax{1.54}

\def\tauBmin{0.01}
\def\tauBmax{0.18}

\def\NAmin{0.2}
\def\NAmax{37.1}
\def\NAorder{16}

\def\NBmin{0.4}
\def\NBmax{35.1}
\def\NBorder{15}

\def\Xmin{5.7}
\def\Xmax{33.0}

\def\Taverage{28.4}
\def\Tstddev{9.7}

\def\Xall{12.3}

\def\XBar{16.58}
\def\XBarerror{0.17}
\def\XDLSF{15.83}
\def\XDLSFerror{0.14}
\def\XOMCfour{11.64}
\def\XOMCfourerror{0.05}
\def\XOMCtwothree{12.92}
\def\XOMCtwothreeerror{0.04}
\def\XBend{11.68}
\def\XBenderror{0.04}
\def\XOMCone{12.14}
\def\XOMConeerror{0.04}

\def\XBarpixel{108}
\def\XDLSFpixel{371}
\def\XOMCfourpixel{2150}
\def\XOMCtwothreepixel{5487}
\def\XBendpixel{5061}
\def\XOMConepixel{2673}

\usepackage{graphicx}
\usepackage{txfonts}
\usepackage{natbib}
\begin{document}

   \title{High abundance ratio of $^{13}$CO to C$^{18}$O toward photon-dominated regions in the Orion-A giant molecular cloud \thanks{Figures \ref{co_peak_map}, \ref{mom0_maps} (a) and (b), and \ref{13co_channel_maps} (a) are available in electric form at the CDS via anonymous ftp to cdsarc.u-strasbg.fr (130.79.128.5) or via http://cdsweb.u-strasbg.fr/cgi-bin/qcat?J/A+A/.}}

   \subtitle{}

   \author{
Yoshito Shimajiri\inst{1,2,3}
          \and  
Yoshimi Kitamura\inst{4}
          \and
Masao Saito\inst{2,5}
          \and
Munetake Momose\inst{6}
          \and
Fumitaka Nakamura\inst{2}
          \and 
Kazuhito Dobashi\inst{7}
          \and 
Tomomi Shimoikura\inst{7}
          \and   
Hiroyuki Nishitani\inst{2,3}
          \and
Akifumi Yamabi\inst{7}
          \and 
Chihomi Hara\inst{2,8}
          \and 
Sho Katakura\inst{7}
          \and 
Takashi Tsukagoshi\inst{6} 
          \and
Tomohiro Tanaka\inst{9}
          \and 
Ryohei Kawabe\inst{2,5}
}

   \institute{
Laboratoire AIM, CEA/DSM-CNRS-Universit$\acute{e}$ Paris Diderot, IRFU/Service d'Astrophysique, CEA Saclay, F-91191 Gif-sur-Yvette, France
              \email{Yoshito.Shimajiri@cea.fr}
         \and
National Astronomical Observatory of Japan, 2-21-1 Osawa, Mitaka, Tokyo 181-8588, Japan
         \and
Nobeyama Radio Observatory, 462-2 Nobeyama, Minamimaki, Minamisaku, Nagano 384-1305, Japan
         \and
Institute of Space and Astronautical Science, Japan Aerospace Exploration Agency, 3-1-1 Yoshinodai, Chuo-ku, Sagamihara 252-5210, Japan
         \and
Joint ALMA Observatory, Alonso de Cordova 3107 Vitacura, Santiago 763 0355, Chile
                  \and
Institute of Astrophysics and Planetary Sciences, Ibaraki University, 2-1-1 Bunkyo, Mito, Ibaraki 310-8512, Japan
         \and
Department of Astronomy and Earth Sciences, Tokyo Gakugei University, Koganei, Tokyo 184-8501, Japan
         \and
The University of Tokyo, 7-3-1 Hongo Bunkyo, Tokyo 113-0033, Japan
         \and
Department of Physical Science, Osaka Prefecture University, Gakuen 1-1, Sakai, Osaka 599-8531, Japan
             }

   \date{Received October 25 2013; accepted February 24 2014 }

 
  \abstract
   {}
   {We derive physical properties such as the optical depths and the column densities of $^{13}$CO and C$^{18}$O to investigate the relationship between the far ultraviolet (FUV) radiation and the abundance ratios between $^{13}$CO and C$^{18}$O.}
   {We have carried out wide-field (0.4 deg$^2$) observations with an angular resolution of 25.8$\arcsec$ ($\sim$ 0.05 pc) in $^{13}$CO ($J$=1--0) and C$^{18}$O ($J$=1--0) toward the Orion-A giant molecular cloud using the Nobeyama 45 m telescope in the on-the-fly mode.}
   {Overall distributions and velocity structures of the $^{13}$CO and C$^{18}$O emissions are similar to 
those of the $^{12}$CO ($J$=1--0) emission. The optical depths of the $^{13}$CO and C$^{18}$O emission lines are estimated to be \tauAmin\ $<$ $\tau_{\rm ^{13}CO}$ $<$ \tauAmax\ and \tauBmin\ $<$ $\tau_{\rm C^{18}O}$ $<$ \tauBmax, respectively. The column densities of the $^{13}$CO and C$^{18}$O emission lines are estimated to be \NAmin\ $\times$ 10$^{\NAorder}$ $<$ $N_{\rm ^{13}CO}$ $<$ 3.7 $\times$ 10$^{17}$ cm$^{-2}$ and  \NBmin\ $\times$ 10$^{\NBorder}$ $<$ $N_{\rm C^{18}O}$ $<$ 3.5 $\times$ 10$^{16}$ cm$^{-2}$, respectively. The abundance ratios between $^{13}$CO and C$^{18}$O, $X_{\rm ^{13}CO}$/$X_{\rm C^{18}O}$, are found to be \Xmin -- \Xmax. The mean value of $X_{\rm ^{13}CO}$/$X_{\rm C^{18}O}$ in the nearly edge-on photon-dominated regions is found to be 16.47 $\pm$ 0.10, which is a third larger than that the solar system value of 5.5. The mean value of  $X_{\rm ^{13}CO}$/$X_{\rm C^{18}O}$ in the other regions is found to be 12.29 $\pm$ 0.02. The difference of the abundance ratio is most likely due to the selective FUV photodissociation of C$^{18}$O.
}
   {}


   \titlerunning{High abundance ratio of $^{13}$CO and C$^{18}$O}
   \authorrunning{Shimajiri et al.}
\maketitle
%

\section{Introduction}
Far ultraviolet (FUV: 6 eV $<$ $h$$\nu$ $<$ 13.6 eV) radiation emitted from massive stars influences the structure, chemistry, thermal balance, and evolution of the neutral interstellar medium of galaxies \citep{Hollenbach97}. 
Furthermore, stars are formed in the interstellar medium (ISM) irradiated by the FUV radiation. 
Hence, studies of the influence of FUV are crucial for understanding the process of star formation.
Regions where FUV photons dominate the energy balance or chemistry of the gas are called photon-dominated regions (PDRs). 
The FUV emission selectively dissociates CO isotopes more effectively than CO because of the difference in the self shielding \citep{Glassgold85, Yurimoto04, Liszt07, Rollig13}.
The FUV intensity at the wavelengths of the dissociation lines for abundant CO decays rapidly on the surface of molecular clouds, since the FUV emission becomes optically very thick at these wavelengths.
For less abundant C$^{18}$O, which has shifted absorption lines owing to the difference in the vibrational-rotational energy levels, the decay of FUV is much lower. As a result, C$^{18}$O molecules are expected to be selectively dissociated by UV photons, even in a deep molecular cloud interior. 
In the dark cloud near a young cluster (IC 5146, for example), the ratio of the $^{13}$CO to C$^{18}$O fractional abundance, $X_{\rm ^{13}CO}$/$X_{\rm C^{18}O}$, considerably exceeds the solar system value of 5.5 at visual extinction ($A_V$) values of less than 10 mag \citep[Fig. 19 in][]{Lada94}.
This trend indicates the selective UV photodissociation of C$^{18}$O.
A variation of the abundance ratios of the isotopes is also reported \citep{Wilson99,Wang09}. For example, the values of $X_{\rm ^{13}CO}$/$X_{\rm C^{18}O}$ in the solar system, local ISM, Galactic center, and Large Magellanic
Cloud (LMC) are measured to be 5.5, 6.1, 12.5, and 40.8. In the Milky way, the isotopic ratio is proportional to the distance from our Galactic center \citep{Wilson99}.

The Orion-A giant molecular cloud (Orion-A GMC) is the nearest GMC ($d$ = 400 pc; \cite{Menten07,Sandstrom07,Hirota08})  and is one of the best studied star-forming regions (e.g., \citet{Bally87, Dutrey93, Tatematsu99,Johnstone99,Shimajiri08,Shimajiri09, Takahashi08,Davis09,Berne10,Takahashi13,Lee13}). In the northern part of the Orion-A GMC, there are three HII regions, M 42, M 43, and NGC 1977 \citep{Goudis82}. 
From the comparison of the AzTEC 1.1 mm and the Nobeyama 45 m $^{12}$CO ($J$=1--0), and $Midcourse\ Space\ Experiment$ ($MSX$) 8 $\mu$m emissions (from polycyclic aromatic hydrocarbons, PAHs) maps, \citet{Shimajiri11} have identified seven PDRs and their candidates in the northern part of the Orion-A GMC: 1) Orion Bar; 2) the M 43 Shell; 3) a dark lane south filament (DLSF); 4-7) the four regions A-D. 
Since the stratification among these distributions can be recognized, the PDR candidates are likely to be influenced by the FUV emission from the Trapezium star cluster and from NU Ori in nearly edge-on configuration.
Thus, the Orion-A GMC is one of the most suitable targets for investigating the PDRs.
Recently, \citet{Shimajiri13} carried out wide-field (0.17 deg$^2$) and high-angular resolution (21.3$\arcsec$ $\sim$ 0.04 pc) observations in [CI] line toward the Orion-A GMC. 
The mapping region includes the nearly edge-on PDRs and the four PDR candidates of the Orion Bar, DLSF, M 43 Shell, and Region D. 
The overall distribution of the [CI] emission coincides with that of the $^{12}$CO emission in the nearly edge-on PDRs, which is inconsistent with the prediction by the plane-parallel PDR model \citep{Hollenbach99}. The [CI] distribution in the Orion-A GMC is found to be more similar to those of the $^{13}$CO ($J$=1--0), C$^{18}$O ($J$=1--0), and H$^{13}$CO$^+$ ($J$=1--0) lines rather than that of the $^{12}$CO ($J$=1--0) line in the inner part of the cloud, suggesting that the [CI] emission is not limited to the cloud surface, but is tracing the dense, inner parts of the cloud.

\begin{table*}
\centering
\caption{Parameters of our observations \label{observations_parameter}}
\begin{tabular}{lcc}
\hline
\hline
Molecular line          & $^{13}$CO ($J$=1--0) & C$^{18}$O ($J$=1--0)  \\
 \hline
Rest Frequency [GHz] & 110.201354 & 109.782176\\
Observation & 2013 May & 2010 March -- 2013 May\\
Scan mode & OTF & OTF \\
Mapping size [deg$^{2}$] & 0.4  & 0.4 \\
Effective beam size [$\arcsec$] & 25.8 & 25.8 \\
Velocity resolution [km s$^{-1}$] & 0.3  & 0.3  \\
Typical noise level in $T_{\rm MB}$ [K] & 0.7   & 0.2  \\
\hline
\end{tabular}
\end{table*}

\begin{figure}
\begin{center}
\includegraphics[width=90mm, angle=0]{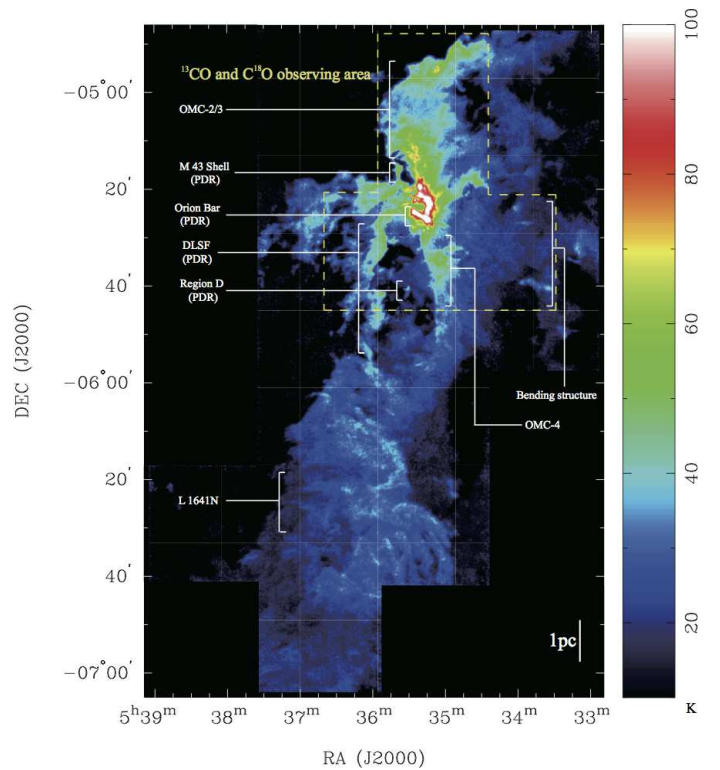}
\caption{Peak intensity map in the $^{12}$CO ($J$=1--0) line in units of K ($T_{\rm MB}$). 
The data are from \citet{Shimajiri11} and \citet{Nakamura12}. A dashed box shows the $^{13}$CO ($J$=1--0) and C$^{18}$O ($J$=1--0)  observing region. 
The $^{12}$CO data in FITS format are available at the NRO web page via http://www.nro.nao.ac.jp/~nro45mrt/html/results/data.html.}
\label{co_peak_map}
\end{center}
\end{figure}

This paper is organized as follows: In Sect. 2, the Nobeyama 45 m observations are described. In Sect. 3, we present the $^{13}$CO and C$^{18}$O maps of the Orion-A GMC and estimate the optical depths of the $^{13}$CO and C$^{18}$O gas and the column densities of these molecules. In Sect. 4, we discuss the variation of the ratio of the $^{13}$CO to C$^{18}$O fractional abundance in terms of the FUV radiation. In Sect. 5, we summarize our results. Detailed distributions of the filaments and dense cores and their velocity structure and mass will be reported in a forthcoming paper.

\section{NRO 45m observations \& data reduction}

In 2010 and 2013, we carried out $^{13}$CO ($J$=1--0) and C$^{18}$O ($J$=1--0) mapping observations toward a 0.4 deg$^2$ region in the northern part of the Orion-A GMC with the 25-element focal plane receiver BEARS installed in the 45 m telescope at the Nobeyama Radio Observatory (NRO). The $^{13}$CO ($J$=1--0) and C$^{18}$O ($J$=1--0) data were obtained separately.
Figure \ref{co_peak_map} shows the $^{13}$CO and C$^{18}$O observing areas. At 110 GHz, the telescope has a beam size of 16$\arcsec$ (HPBW) and a main beam efficiency, $\eta_{\rm MB}$, of 38 \% in the 2010 season and 36 \% in 2013, which are from observatory measurements at 110 GHz using the S100 receiver.
The beam separation of the BEARS is 41$\arcsec$.1 on the sky plane \citep{Sunada00}.  As the back end, we used 25 sets of 1024 channel auto-correlators (ACs) which have a 32 MHz bandwidth and a frequency resolution of 37.8 kHz \citep{Sorai00}. The frequency resolution corresponds to a velocity resolution of $\sim$ 0.1 km s$^{-1}$ at 110 GHz. During the observations, the system noise temperatures were in a range from 270 K to 470 K in the double sideband (DSB). 
The standard chopper wheel method was used to convert the observed signal to the antenna temperature, $T_{\rm A}^{*}$, in units of K, corrected for the atmospheric attenuation. 
The data are given in terms of the main-beam brightness temperature, $T_{\rm MB} = T_{\rm A}^*/\eta_{\rm MB}$. 
The telescope pointing was checked every 1.5 hours by observing the SiO maser source Ori KL, and was better than 3$\arcsec$ throughout the entire observation. 
The intensity scales of the BEARS 25 beams are different from each other owing to the varying sideband ratios of the beams since the BEARS receiver is operated in DSB mode. 
To calibrate the different intensity scales, we used calibration data obtained from the observations toward W3 and Orion IRC 2 using another SIS receiver, S100, with a single sideband (SSB) filter and acousto-optical spectrometers (AOSs).
The intensity scales between the S100 receiver and the BEARS 25 beams were estimated to be 1.96 -- 3.96 for $^{13}$CO and 1.57 - 2.77 for C$^{18}$O.

Our mapping observations were made with the on-the-fly (OTF) mapping technique \citep{Sawada08}. 
We used an emission-free area $\sim$ 2$^{\circ}$ away from the mapping area as the off positions.
We obtained OTF maps with two different scanning directions along the RA or Dec axes covering the 20$\arcmin$ $\times$ 20$\arcmin$ or the 20$\arcmin$ $\times$ 10$\arcmin$ regions, and combined them into a single map to reduce the scanning effects as much as possible.  
We adopted a spheroidal function$\footnote[1]{\citet{Schwab84} is a convolution kernel and \citet{Sawada08} described the details of the spheroidal function. We applied the parameters $m$ = 6 and $\alpha$ = 1, which define the shape of the function.}$ \citep{Sawada08} to calculate the intensity at each grid point of the final cube data with a spatial grid size of 10$\arcsec$, resulting in the final effective resolution of 25$\arcsec$.8. 
The 1 $\sigma$ noise level of the final data is 0.5 K for $^{13}$CO and 0.14 K for C$^{18}$O in $T_{\rm MB}$ at a velocity resolution of 0.8 km s$^{-1}$. 
By combining scans along the R.A. and Dec. directions, we minimized the so-called scanning effect using the \citet{Emerson88} PLAIT algorithm. 
We summarize the parameters for the $^{13}$CO ($J$=1--0) and C$^{18}$O ($J$=1--0) line observations in Table \ref{observations_parameter}.

\begin{figure*}
\begin{center}
\includegraphics[width=60mm, angle=0]{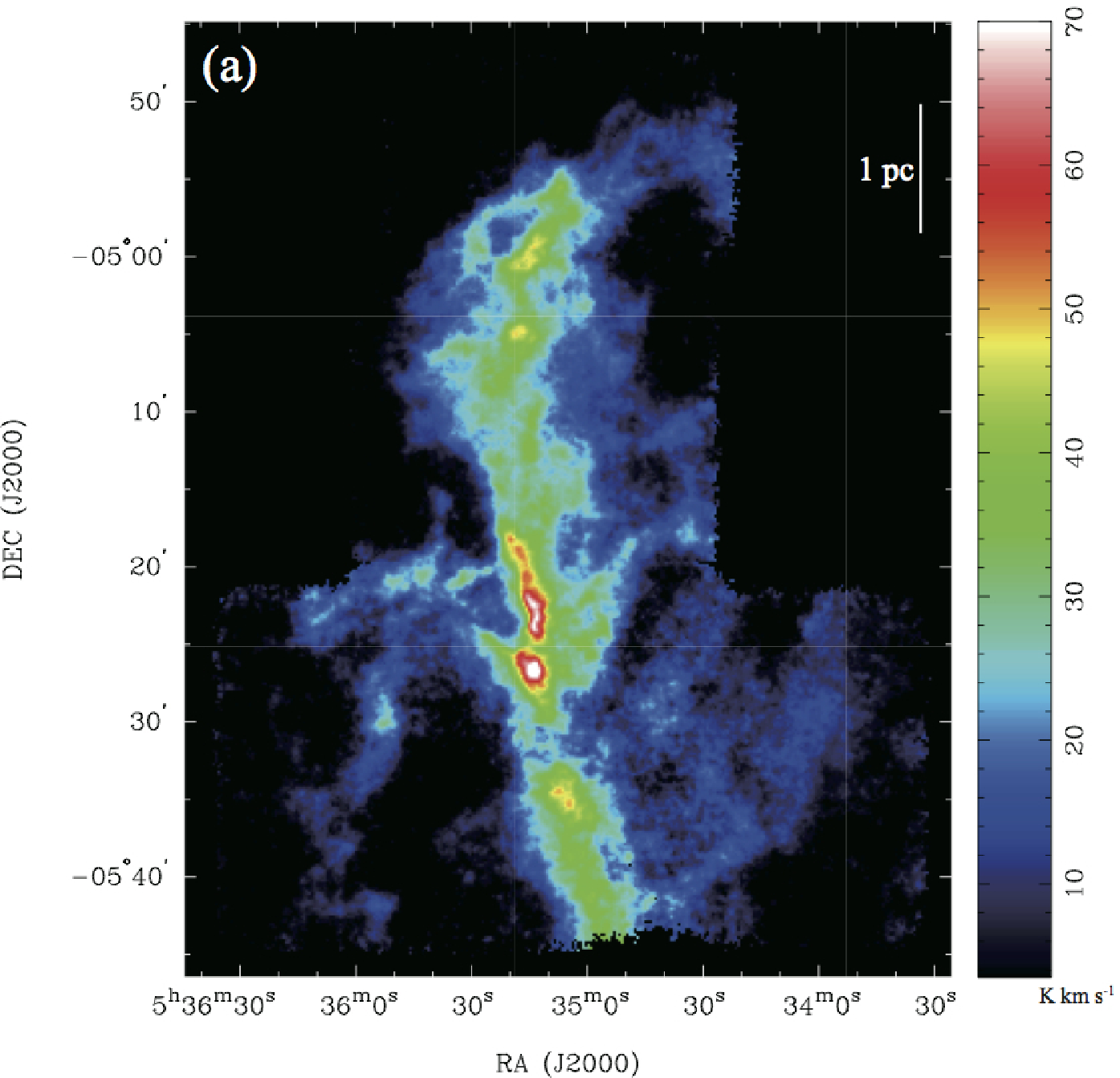}
\includegraphics[width=60mm, angle=0]{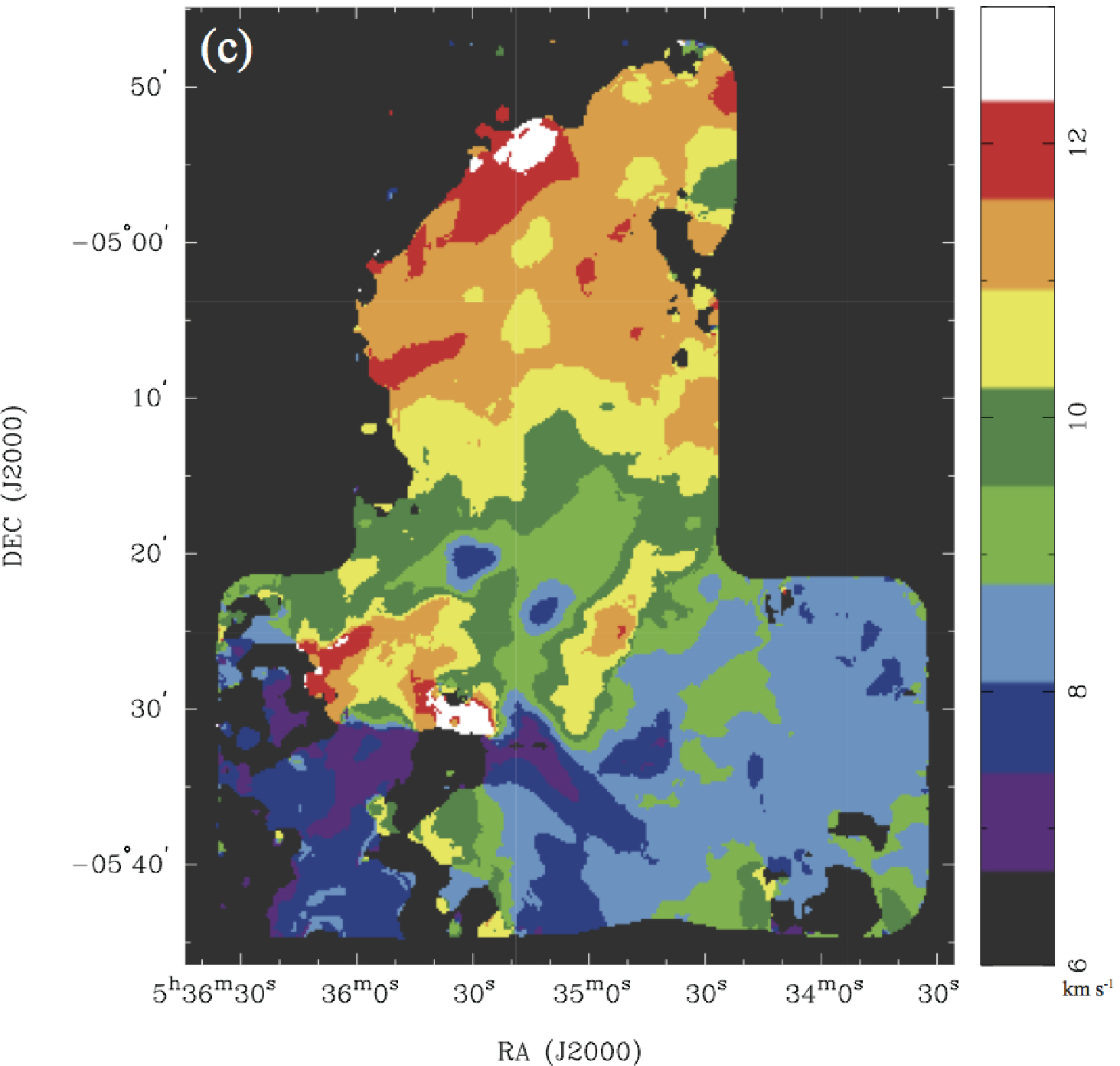}
\includegraphics[width=60mm, angle=0]{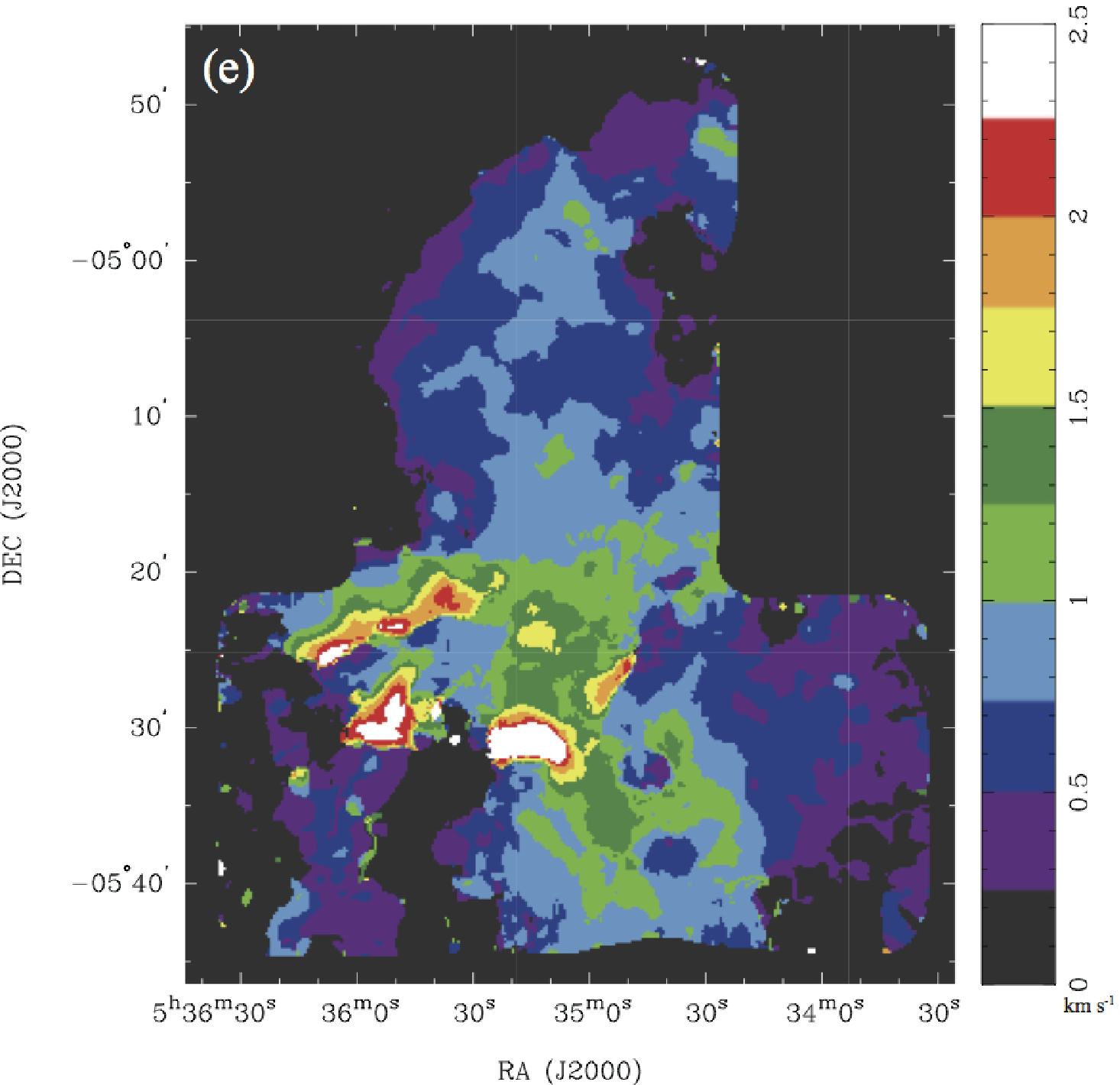}
\includegraphics[width=60mm, angle=0]{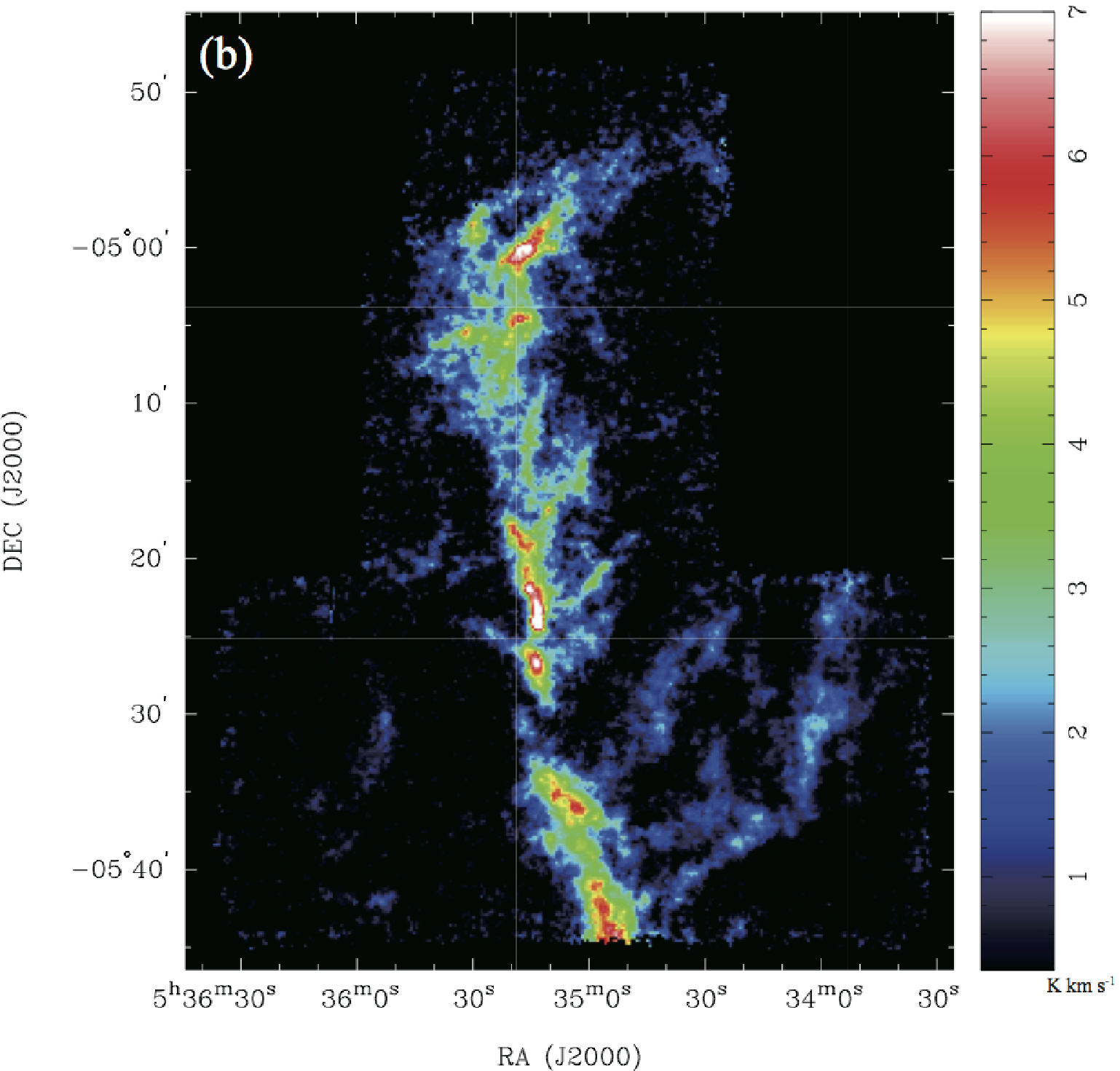}
\includegraphics[width=60mm, angle=0]{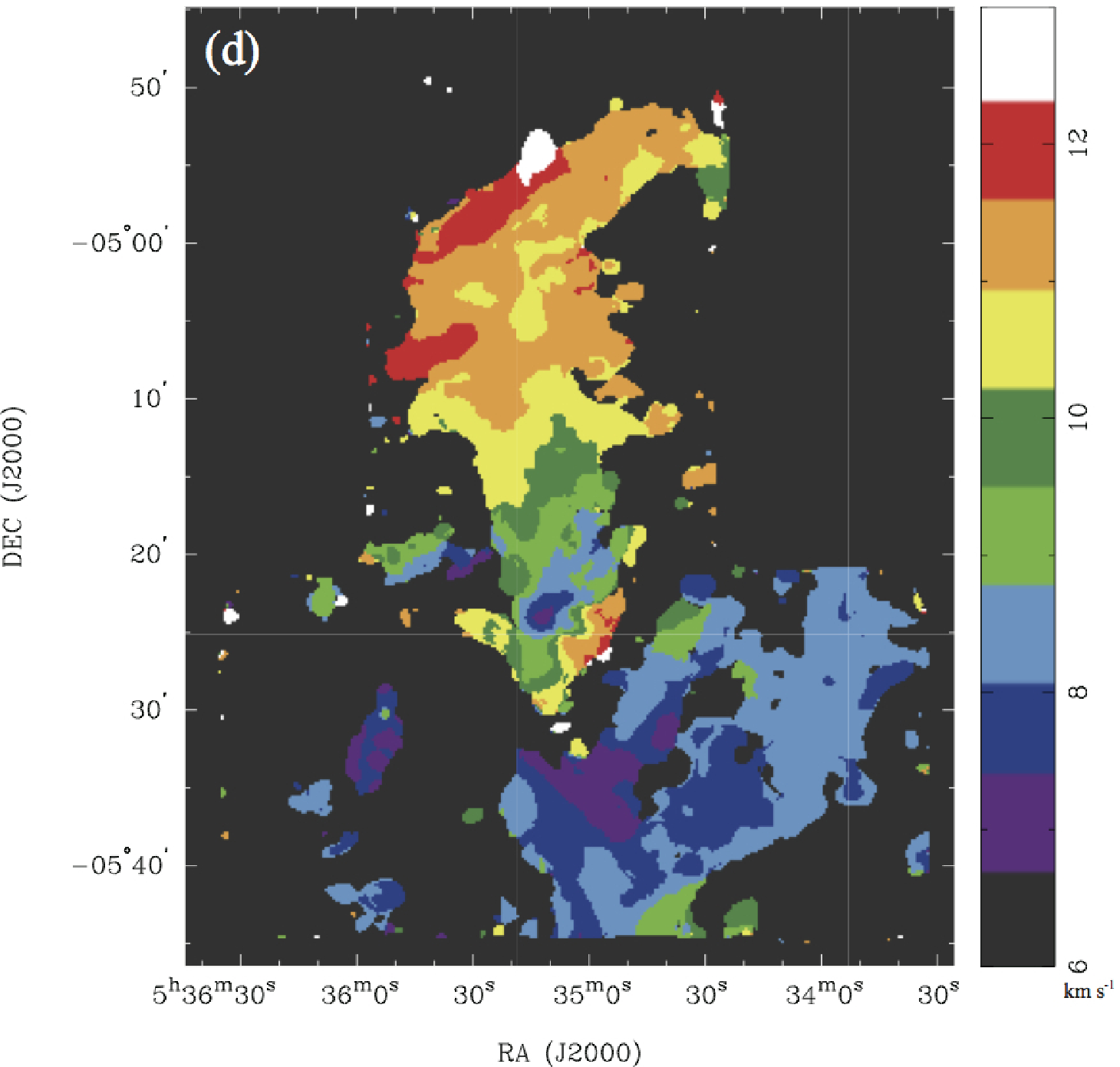}
\includegraphics[width=60mm, angle=0]{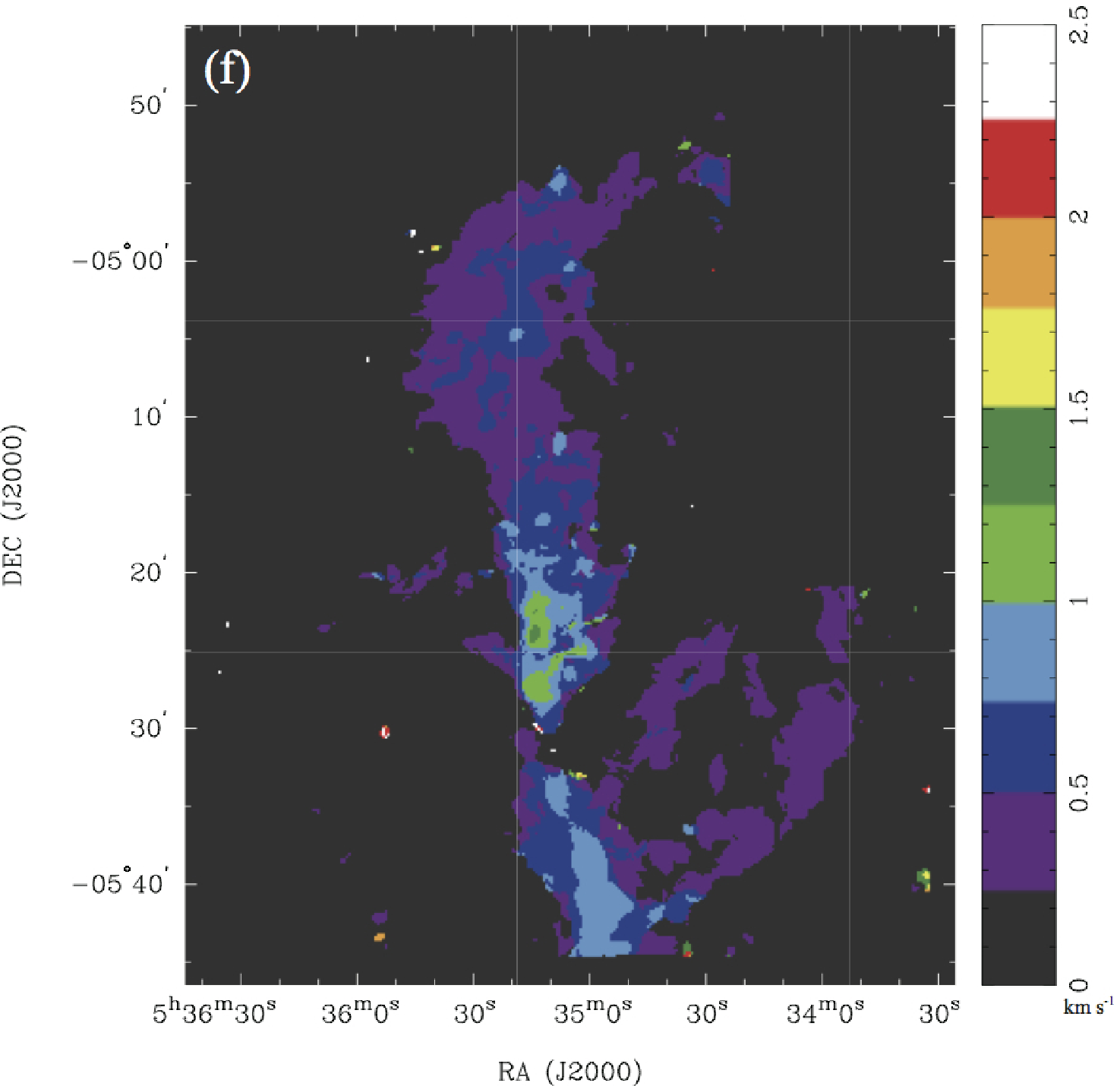}
\caption{Total intensity maps of the (a) $^{13}$CO ($J$=1--0) and (b) C$^{18}$O ($J$=1--0) emission lines
integrated over the velocity range 4.25 $<$ $V_{\rm LSR}$ $<$ 14.25 km s$^{-1}$ in units of K km s$^{-1}$ ($T_{\rm MB}$).
The other panels are the mean velocity maps of (c) $^{13}$CO and (d) C$^{18}$O in units of km s$^{-1}$, 
and the velocity dispersion maps of (e) $^{13}$CO and (f) C$^{18}$O in units of km s$^{-1}$ calculated for the same velocity range for the total intensity maps.
The $^{13}$CO and C$^{18}$O data in FITS format are available at the NRO web page via http://www.nro.nao.ac.jp/~nro45mrt/html/results/data.html.}
\label{mom0_maps}
\label{mom1_maps}
\label{mom2_maps}
\end{center}
\end{figure*}

\section{Results}
\subsection{Distributions of the $^{13}$CO ($J$=1--0) and C$^{18}$O ($J$=1--0) emission lines }

Figures \ref{mom0_maps} (a) and (b) show the total intensity maps of the $^{13}$CO ($J$=1--0) and C$^{18}$O ($J$=1--0) emission lines, respectively, integrated over a velocity range  4.25 $<$ $V_{\rm LSR}$ $<$ 14.25 km s$^{-1}$ in the northern part of the Orion-A GMC. 
The overall distributions of the $^{13}$CO and C$^{18}$O emission lines are found to be similar to that of the $^{12}$CO ($J$=1--0) emission line by \citep{Shimajiri11}. 
In the $^{13}$CO map, the brightest position is at (R.A., Dec.) = (5$^{\rm h}$35$^{\rm m}$13.$^{\rm s}$0, -5$^{\circ}$26$\arcmin$48$\arcsec$.24), which is located on the western side of the Orion bar and is close to the peak position of the [CI] emission \citep{Shimajiri13}. 
On the other hand, the peak position in the C$^{18}$O map is at (R.A., Dec.) = (5$^{\rm h}$35$^{\rm m}$15.$^{\rm s}$7, -5$^{\circ}$22$\arcmin$00$\arcsec$.24) and is different from the brightest position in the $^{13}$CO map. 
The elongated structure along the north-south direction is a part of the integral-shaped filament observed in the dust-continuum emissions at 850 $\mu$m \citep{Johnstone99}, 1.2 mm \citep{Davis09}, and 1.3 mm \citep{Chini97} as well as in molecular lines such as $^{13}$CO ($J$=1--0), C$^{18}$O ($J$=1--0), CS ($J$=1--0) \citep{Tatematsu93}, and  H$^{13}$CO$^+$ ($J$=1--0) \citep{Ikeda07}. 
We note that the size of our mapping area centered at the filament  is $\sim$ 40$\arcmin$ in the east-west direction, wider than those of the previous observations in the $^{13}$CO and C$^{18}$O ($J$=1--0) lines by \citep{Tatematsu93}.  
There is a branch of the $^{13}$CO and C$^{18}$O emissions extending toward the southeast of Ori-KL. 
This structure, called a dark lane south filament, corresponds to one of the PDRs \citep{Rodriguez01, Shimajiri11}. 
Moreover,  a bending filamentary structure with a length of $\sim$ 3 pc, is seen to the southwest of Ori-KL (also see Fig. \ref{X_maps}). These features are also found in our previous observations in the 1.1 mm dust continuum and $^{12}$CO ($J$=1--0) emissions \citep{Shimajiri11}. Hereafter, following \citet{Shimajiri11}, we call these structures the DLSF and the bending structure.

\subsection{Velocity structures of the $^{13}$CO and C$^{18}$O emission lines}

\begin{figure*}
\begin{center}
\includegraphics[width=170mm, angle=0]{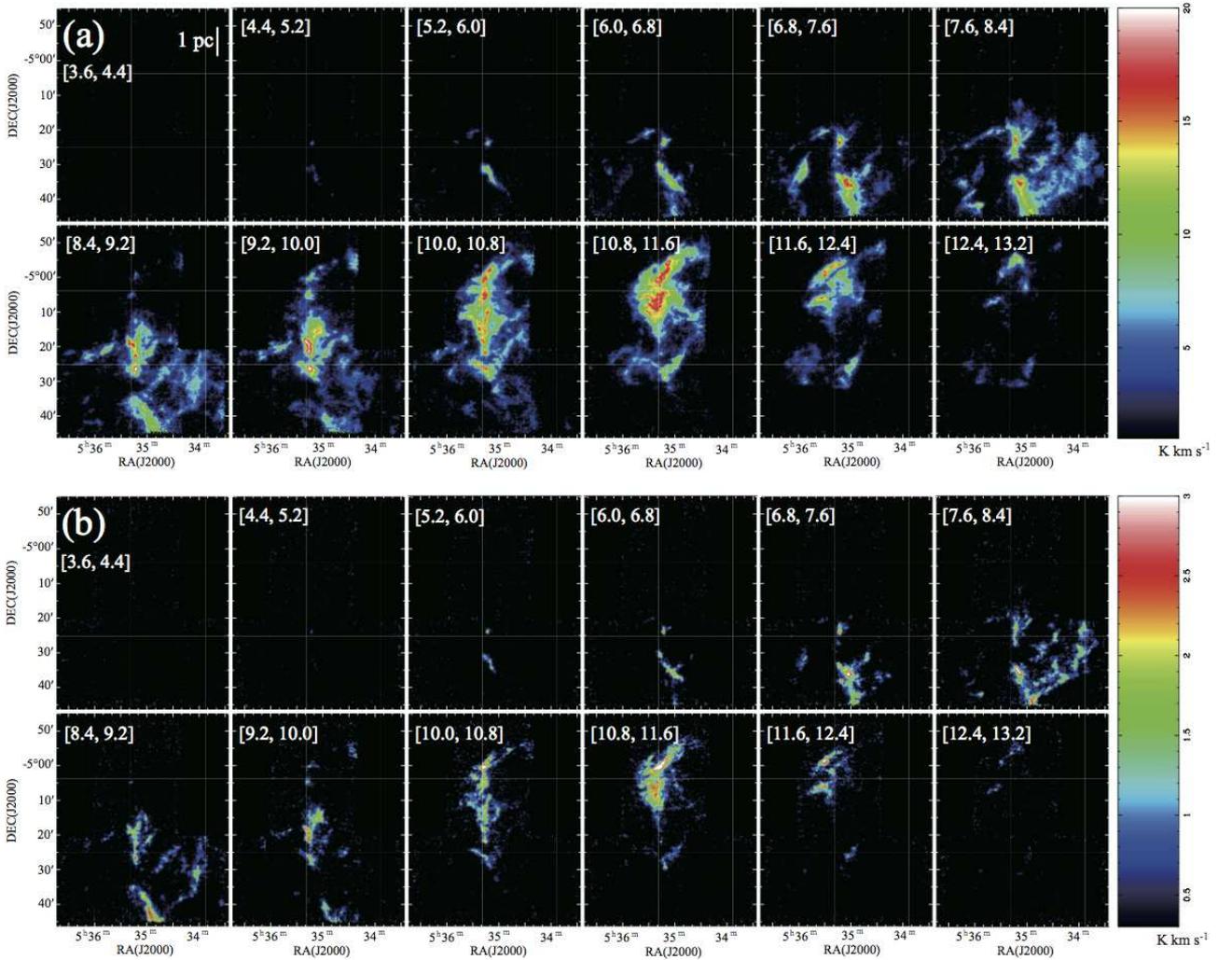}
\caption{
Velocity channel maps of the (a) $^{13}$CO ($J$=1--0) and (b) C$^{18}$O ($J$=1--0) emission line in units of K km s$^{-1}$ ($T_{\rm MB}$ $dV$). The velocity range used for the integration is indicated in the top-left corner of each panel. }
\label{13co_channel_maps}
\label{c18o_channel_maps}
\end{center}
\end{figure*}

Figures \ref{13co_channel_maps} (a) and (b) show the $^{13}$CO ($J$=1--0) and C$^{18}$O ($J$=1--0) velocity channel maps over the range 3.6 $<$ $V_{\rm LSR}$ $<$ 13.2 km s$^{-1}$.
In the maps for 4.4 $<$ $V_{\rm LSR}$ $<$ 9.2 km s$^{-1}$, the $^{13}$CO and C$^{18}$O emission lines are distributed in the southern part of the mapping area. 
On the other hand, 
in the maps for 10.0 $<$ $V_{\rm LSR}$ $<$ 13.2 km s$^{-1}$, the emission lines are seen in the northern part. These results found the large-scale velocity gradient ($\sim$ 1 km s$^{-1}$ pc$^{-1}$ ), which has been seen in the previous studies in the $^{12}$CO, $^{13}$CO, H$^{13}$CO$^{+}$, and CS lines \citep{Bally87, Tatematsu93, Ikeda07, Shimajiri11, Buckle12}. 
The velocity gradient from south to north is also seen in the mean velocity maps\footnote[2]{The mean velocity maps are produced by using the task MOMENT in MIRIAD and are calculated using the equation $\bar{V}=\int T_{\rm MB} V dV/\int T_{\rm MB} dV$. We made them from the $^{13}$CO and C$^{18}$O velocity channel maps with a velocity resolution of 0.3 km s$^{-1}$ with a clip equal to twice the respective map rms noise level. The 1 $\sigma$ noise levels of the $^{13}$CO and C$^{18}$O velocity channel maps are 0.7 K and 0.2 K in units of $T_{\rm MB}$.} of the $^{13}$CO and C$^{18}$O emission lines shown in Figs \ref{mom2_maps} (c) and \ref{mom2_maps} (d), respectively.
In the maps for 10.8 $<$ $V_{\rm LSR}$ $<$ 13.2 km s$^{-1}$, the $^{13}$CO emission line shows a shell-like structure with a size of about two pc toward the south of Ori-KL. 
Its velocity structure does not follow the large-scale velocity gradient from south to north. 
The shell can also be recognized in the $^{12}$CO map (see Fig 5 in \citet{Shimajiri11}.
The overall velocity structures of $^{13}$CO and C$^{18}$O are consistent with those of $^{12}$CO \citep{Shimajiri11}. 
However, in one of the $^{13}$CO channel maps at $V_{\rm LSR}$ =10.0 -- 10.8 km s$^{-1}$, a shell-like feature centered at (R.A., Dec.) = (5$^{\rm h}$34$^{\rm m}$47$^{\rm s}$, -5$^{\circ}$32$\arcmin$24$\arcsec$) can be recognized, and this feature is not obvious in the $^{12}$CO and C$^{18}$O maps.

\begin{table}
\centering
\caption{Comparison of velocity dispersion between Orion-A GMC and L1551 \label{velocity_dispersion}}
\begin{tabular}{lcc}
\hline
\hline
Molecular line           & Orion-A GMC &  L 1551 \\
 \hline
$^{13}$CO ($J$=1--0) & 0.67 $\pm$ 0.34 km s$^{-1}$ & 0.22 $\pm$ 0.09 km s$^{-1}$ \\
C$^{18}$O ($J$=1--0) & 0.53 $\pm$ 0.21 km s$^{-1}$ & 0.13 $\pm$ 0.05 km s$^{-1}$ \\
\hline
\end{tabular}
\end{table}

Figures \ref{mom2_maps} (e) and (f) show the velocity dispersion maps\footnote[3]{The velocity dispersion maps are produced by using the task MOMENT in MIRIAD and are calculated using the equation $\sigma^2=\int T_{\rm MB} (V-\bar{V}) dV/\int T_{\rm MB} dV$. We made them from the $^{13}$CO and C$^{18}$O velocity channel maps with a velocity resolution of 0.3 km s$^{-1}$ with a clip equal to twice the respective map rms noise level.} of the $^{13}$CO ($J$=1--0) and C$^{18}$O ($J$=1--0) emission lines calculated over the velocity range 5.2 $<$ $V_{\rm LSR}$ $<$ 14.0 km s$^{-1}$.
The mean values of the $^{13}$CO and C$^{18}$O velocity dispersions in the observed area are 0.67 $\pm$ 0.34 km s$^{-1}$ (min: 0.30 km s$^{-1}$, max: 3.6 km s$^{-1}$) and 0.53 $\pm$ 0.21 km s$^{-1}$ (min: 0.30 km s$^{-1}$, max: 2.8 km s$^{-1}$), respectively. 
In Table 2, we compare the velocity dispersions in the Orion-A GMC and a low-mass star-forming region, L 1551 \citep{Yoshida10}.
The velocity dispersions in the Orion-A GMC are three to four times larger than those in L 1551. 
Both the $^{13}$CO and C$^{18}$O velocity dispersion maps show that the velocity dispersion becomes the largest in the OMC 1 region.
In addition, the velocity dispersion in the integral-shaped filament is relatively high compared with that in the outer region of the filament. 
The $^{12}$CO velocity dispersion increases toward the east of the OMC-2/3 filament \citep[see Fig. 7 in ][]{Shimajiri11}. 
However, these increments in the $^{13}$CO and C$^{18}$O velocity dispersions are not recognized in our maps.

\begin{figure*}
\begin{center}
\includegraphics[width=180mm, angle=0]{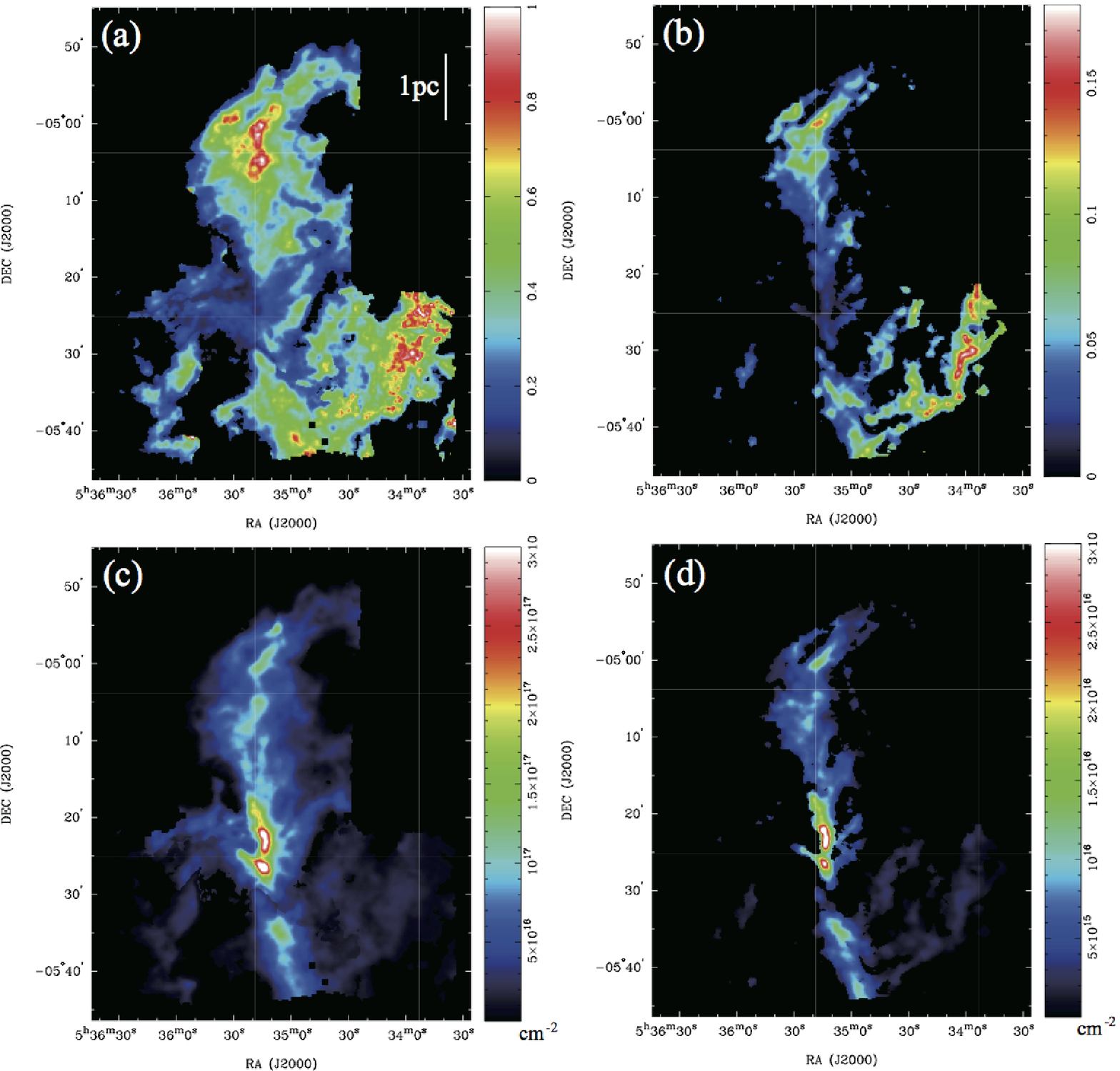}
\caption{
Maps of the optical depths of the (a) $^{13}$CO ($J$=1--0) and (b) C$^{18}$O ($J$=1--0) emission lines,
and maps of the column densities of (c) $^{13}$CO and (d) C$^{18}$O molecules. 
The optical depths and column densities are estimated on the assumption that the excitation temperatures of the $^{13}$CO and C$^{18}$O ($J$=1--0) lines are equal to the peak temperatures of the $^{12}$CO ($J$=1--0) emission in $T_{\rm MB}$ and are calculated for pixels with intensities above the 8 $\sigma$ noise levels.
}
\label{tau_maps}
\label{N_maps}
\end{center}
\end{figure*}

\subsection{Column densities of the $^{13}$CO and C$^{18}$O gas and the abundance ratio of $^{13}$CO to C$^{18}$O} \label{depth}

\begin{figure*}
\begin{center}
\includegraphics[width=180mm, angle=0]{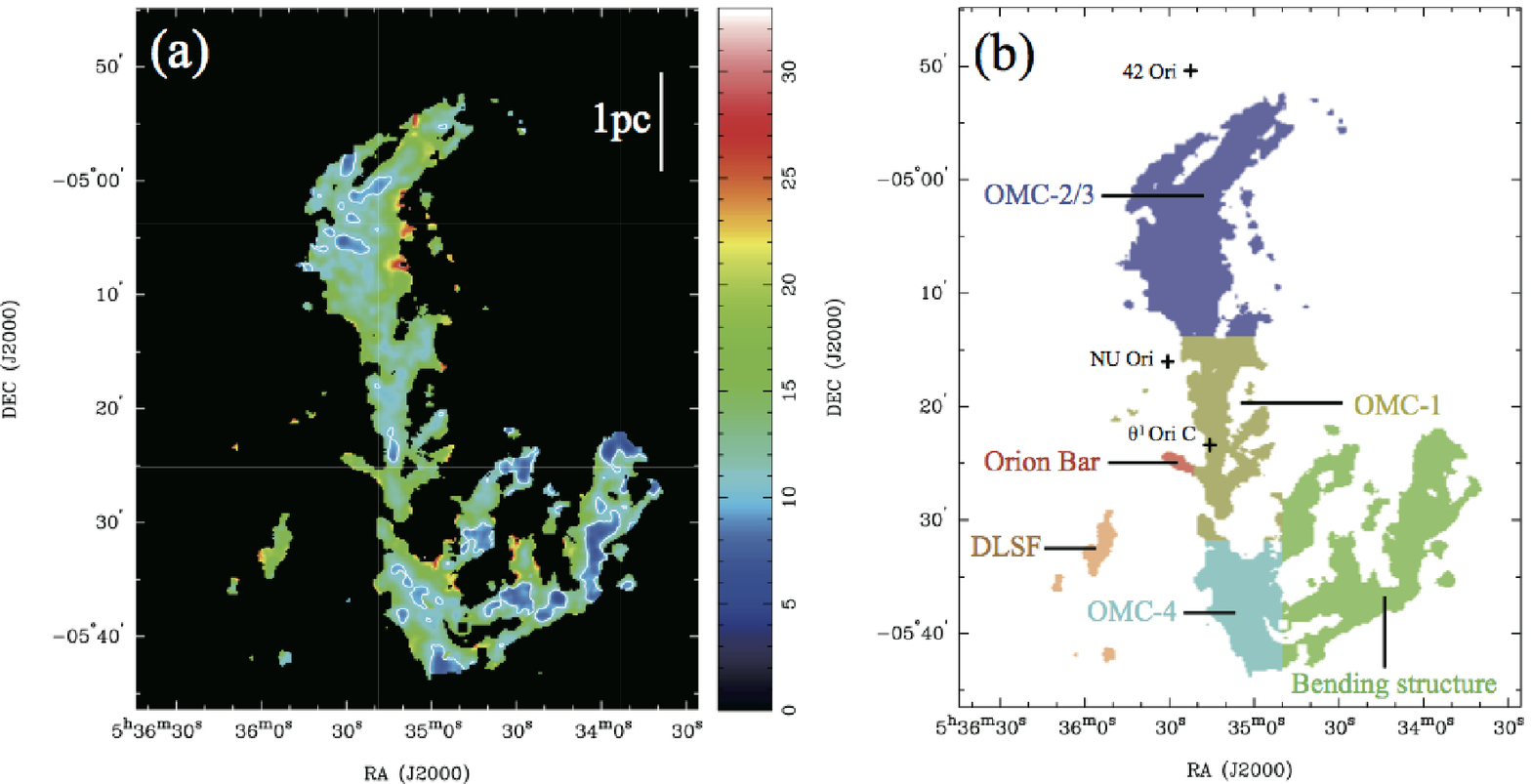}
\caption{
(a) Map of the abundance ratio $X_{\rm ^{13}CO}$/$X_{\rm C^{18}O}$.
(b) Locations of the regions summarized in Table \ref{ratio_of_each_area}. Colors to indicate the individual regions are the same as those of the plots in Fig. \ref{correlation}. 
In the panel (a), the contours show the value of $X_{\rm ^{13}CO}$/$X_{\rm C^{18}O}$ = 10. 
The crosses show
the positions of $\theta^1$ Ori C, NU Ori, and 42 Ori, which are the exciting stars of the HII regions, M 42, M 43, and NGC 1977. 
}
\label{X_maps}
\end{center}
\end{figure*}
Previous studies in the Orion-A GMC revealed that the mean density values, $\bar{n}$, of the filaments traced in $^{13}$CO and the dense cores traced in C$^{18}$O are $\sim$2 $\times$ 10$^3$ cm$^{-3}$ and $\sim$ 5 $\times$ 10$^3$ cm$^{-3}$, respectively, assuming a cylinder and sphere \citep{Nagahama98, Ikeda09}.
These values are comparable to the critical densities of the $^{13}$CO and C$^{18}$O ($J$=1--0) lines, validating the local thermodynamic equilibrium (LTE) assumption.
We estimated the column densities of $^{13}$CO and C$^{18}$O on the assumption that the rotational levels of the $^{13}$CO and C$^{18}$O gas are in the LTE. 
The optical depths and column densities of these molecules, $\tau_{\rm X}$ and $N_{\rm X}$ where X=$^{13}$CO and C$^{18}$O, can be derived using the equations (e.g., \citet{Kawamura98})

\begin{equation}\label{tau13co}
\centering
\tau_{\rm ^{13}CO} = - {\rm ln} \left\{ 1 -\frac{T_{\rm MB}({\rm ^{13}CO)}/\phi_{\rm ^{13}CO}}{5.29[J(T_{\rm ex})-0.164] }  \right\},
\end{equation}

\begin{equation}\label{N13co}
\centering
N_{\rm ^{13}CO} = 2.42 \times 10^{14} \left\{ \frac{\tau_{\rm ^{13}CO} \Delta V({\rm ^{13}CO}) T_{\rm ex}}{1-{\rm exp}[-5.29/T_{\rm ex}]} \right\} \ {\rm cm}^{-2},
\end{equation}

\begin{equation}\label{tauc18o}
\centering
\tau_{\rm C^{18}O} = - {\rm ln} \left\{ 1 -\frac{T_{\rm MB}({\rm C^{18}O)}/\phi_{\rm C^{18}O}}{5.27[J(T_{\rm ex})-0.1666] }  \right\},
\end{equation}

\noindent and

\begin{equation}\label{Nc18o}
\centering
N_{\rm C^{18}O} = 2.42 \times 10^{14} \left\{ \frac{\tau_{\rm C^{18}O} \Delta V({\rm C^{18}O}) T_{\rm ex}}{1-{\rm exp}[-5.27/T_{\rm ex}]} \right\} \ {\rm cm}^{-2},
\end{equation}

\noindent where $T_{\rm ex}$ is the excitation temperature of these molecules in K, and $J[T]$ = 1 / (exp[5.29/$T$] -1) for $^{13}$CO and $J[T]$ = 1/ (exp[5.27/$T$] -1) for C$^{18}$O. 
The beam filling factors of $\phi_{\rm ^{13}CO}$ and $\phi_{\rm C^{18}O}$ for $^{13}$CO and C$^{18}$O, respectively, are assumed to be 1.0 (also see Sect. \ref{filling_section}). 
Here, Eqs. (\ref{N13co}) and (\ref{Nc18o}) assume that the line profile can be approximated by a Gaussian function. 
The peak brightness temperature, $T_{\rm MB}$, and the FWHM line widths, $\Delta V$ ($^{13}$CO) and $\Delta V$ (C$^{18}$O) are in units of K and km s$^{-1}$, respectively.  
We derived $T_{\rm MB}$ and $\Delta V$ by fitting a Gaussian to the observed spectrum at each pixel.
We considered the peak brightness temperature of $^{12}$CO ($J$=1--0), $T_{\rm CO\ peak}$, at each position (Fig. \ref{co_peak_map}) as the $T_{\rm ex}$ values of $^{13}$CO and C$^{18}$O, assuming that the $^{12}$CO emission is optically thick and that the $^{12}$CO lines are not self-absorbed by cold foreground gas. The range of $T_{\rm CO\ peak}$ is from 12.7 K to 108.0 K and the mean value is 35.0 K.

For the pixels having signal-to-noise ratios greater than 8, we derived the optical depths and the column densities as shown in Fig. \ref{N_maps}. 
The optical depths of the $^{13}$CO and C$^{18}$O lines are estimated to be  \tauAmin\ $<$ $\tau_{\rm ^{13}CO}$ $<$ \tauAmax\ and \tauBmin\ $<$ $\tau_{\rm C^{18}O}$ $<$ \tauBmax, respectively (see Figs. \ref{N_maps} (a) and (b)). 
The C$^{18}$O emission is optically thin for most of the observed region. 
Although the column density of the Orion-A GMC is high, molecules in the $J$=1-0 transition lines  become optically thin owing to the high temperature of $>$ 20 K.
In the northern and southern parts of the mapping area where the temperature is relatively low, the optical depths are relatively thick. In contrast, in the central part where the temperature is large, the optical depth becomes small. 
The column densities of the $^{13}$CO and C$^{18}$O gas are estimated to be \NAmin\ $\times$ 10$^{\NAorder}$ $<$ $N_{\rm ^{13}CO}$ $<$ 3.7 $\times$ 10$^{17}$\ and  \NBmin\ $\times$ 10$^{\NBorder}$ $<$ $N_{\rm C^{18}O}$ $<$ 3.5 $\times$ 10$^{16}$ cm$^{-2}$, respectively.  
The column density around Ori-KL is high owing to the high temperature (above 100 K) in spite of the small optical depth. In contrast, the column density in the northern and southern parts is relatively low owing to the low temperature (below 50 K) in spite of the relatively large optical depth.
We summarize these values in Table \ref{Physical_properties}.

Because the fractional abundances of $^{13}$CO ($X_{\rm ^{13}CO}$=$N_{\rm ^{13}CO}$/$N_{\rm H_2}$) and C$^{18}$O ($X_{\rm C^{18}O}$=$N_{\rm C^{18}O}$/$N_{\rm H_2}$) are proportional to their column densities, their abundance ratio can be derived as $X_{\rm ^{13}CO}$/$X_{\rm C^{18}O}$=$N_{\rm ^{13}CO}$/$N_{\rm C^{18}O}$. Based on the observed column densities, we found that the abundance ratio varies in the range of \Xmin\ $<$ $X_{\rm ^{13}CO}$/$X_{\rm C^{18}O}$ $<$ \Xmax\ within the observed area. 
The $X_{\rm ^{13}CO}$/$X_{\rm C^{18}O}$ distribution in Fig. \ref{X_maps} (a) clearly shows that the ratio becomes higher in the nearly edge-on PDRs and the outskirts of the cloud.
In the OMC-1, OMC-2/3, bending structure, and OMC-4 regions, the mean abundance ratio is 12.29 $\pm$ 0.02 (see also Fig. \ref{X_maps} (b)). 
On the other hand, in the nearly edge-on PDRs of the Orion-Bar, and the DLSF, and the outskirts of the cloud, the abundance ratio exceeds 15. The abundance ratio in each region is summarized in Table \ref{ratio_of_each_area}.

\begin{table*}
\centering
\caption{Column densities of the $^{13}$CO and C$^{18}$O gas \label{Physical_properties}}
\begin{tabular}{lcccc}
\hline
\hline
Molecule   &  $T_{\rm ex}$   &  beam filling factor       & Optical depth & Column density \\
 \hline
$^{13}$CO  &   $T_{\rm CO\ peak}$ & 1.0    & \tauAmin -- \tauAmax  &  (\NAmin -- \NAmax) $\times$ 10$^{\NAorder}$ cm$^{-2}$ \\
C$^{18}$O  &   $T_{\rm CO\ peak}$ & 1.0   & \tauBmin -- \tauBmax  &   (\NBmin -- \NBmax) $\times$ 10$^{\NBorder}$ cm$^{-2}$ \\
\hline
$^{13}$CO  &   $T_{\rm CO\ peak}$ &  0.8   &  0.06 -- 4.09 & (0.3 -- 48.0)  $\times$ 10$^{16}$ cm$^{-2}$ \\
C$^{18}$O  &   $T_{\rm CO\ peak}$ &  0.8   &  0.01 -- 0.23 & (0.5 -- 43.9) $\times$ 10$^{15}$   cm$^{-2}$ \\
\hline
$^{13}$CO  &   30 K &  1.0    & 0.05 -- 1.53  & (0.3 -- 22.1) $\times$ 10$^{16}$ cm$^{-2}$ \\
C$^{18}$O  &   20 K &  1.0    & 0.02 -- 0.29  & (0.5 -- 9.9) $\times$ 10$^{15}$ cm$^{-2}$ \\
\hline
$^{13}$CO  &   30 K &  0.8    & 0.06 -- 3.91  &  (0.4 -- 55.0) $\times$ 10$^{16}$cm$^{-2}$ \\
C$^{18}$O  &   20 K &  0.8    & 0.03 -- 0.38  &  (0.6 -- 12.7) $\times$ 10$^{15}$ cm$^{-2}$ \\
\hline

\hline
\end{tabular}
\end{table*}

\begin{table*}
\footnotesize
\centering
\caption{Abundance ratio $X_{\rm ^{13}CO}$/$X_{\rm C^{18}O}$ in the six distinct regions \label{ratio_of_each_area}}
\begin{tabular}{lccccccccc}
\hline
\hline
$T_{\rm ex\ ^{13}CO}$   & $T_{\rm ex\ C^{18}O}$ & $\phi_{\rm ^{13}CO}$ & $\phi_{\rm C^{18}O}$  & Orion-Bar & DLSF & OMC-1 & OMC-2/3 & OMC-4 & Bending structure \\
\hline
$T_{\rm CO\ peak}$    &  $T_{\rm CO\ peak}$  & 1.0                 &   1.0                &  \XBar\ $\pm$ \XBarerror &  \XDLSF\ $\pm$ \XDLSFerror  & \XOMCone\ $\pm$ \XOMConeerror  & \XOMCtwothree\ $\pm$ \XOMCtwothreeerror &  \XOMCfour\ $\pm$ \XOMCfourerror &  \XBend\ $\pm$ \XBenderror \\
$T_{\rm CO\ peak}$    &  $T_{\rm CO\ peak}$  & 1.0                 &   0.8                &  13.25  $\pm$ 0.14 &  12.61  $\pm$ 0.12 & 9.68 $\pm$ 0.03 & 10.26  $\pm$ 0.03 & 9.25  $\pm$ 0.04 & 9.25 $\pm$ 0.03 \\
$T_{\rm CO\ peak}$    &  $T_{\rm CO\ peak}$  & 0.8                 &   0.8                &  16.90 $\pm$ 0.18 & 16.80 $\pm$ 0.16 & 12.53 $\pm$ 0.04 & 14.06 $\pm$ 0.04  & 12.46 $\pm$ 0.05 & 12.61 $\pm$ 0.04  \\
30 K    &  20 K  & 1.0                 &   1.0                &  27.45 $\pm$ 0.25 & 22.82 $\pm$ 0.21 & 20.23 $\pm$ 0.07 & 18.78 $\pm$ 0.06 & 16.25 $\pm$ 0.08 & 14.12  $\pm$ 0.05\\
30 K    &  20 K  & 1.0                 &   0.8                &  21.79  $\pm$ 0.20 & 18.09 $\pm$ 0.17 & 15.99 $\pm$ 0.05 & 14.76 $\pm$ 0.05 & 12.81 $\pm$ 0.06 & 11.18 $\pm$ 0.04\\
30 K    &  20 K  & 0.8                 &   0.8                & 30.27 $\pm$ 0.28  & 25.14 $\pm$ 0.25 & 24.31 $\pm$ 0.11  & 22.17 $\pm$ 0.07 & 18.08 $\pm$ 0.10 & 14.59 $\pm$ 0.06  \\
\hline
Number of pixels     &   &  &  &  \XBarpixel  & \XDLSFpixel  & \XOMConepixel & \XOMCtwothreepixel & \XOMCfourpixel &  \XBendpixel  \\
\hline
\end{tabular}
\end{table*}


\section{Discussion}
\subsection{Selective FUV photodissociation of C$^{18}$O}
\begin{figure*}
\begin{center}
\includegraphics[width=170mm, angle=0]{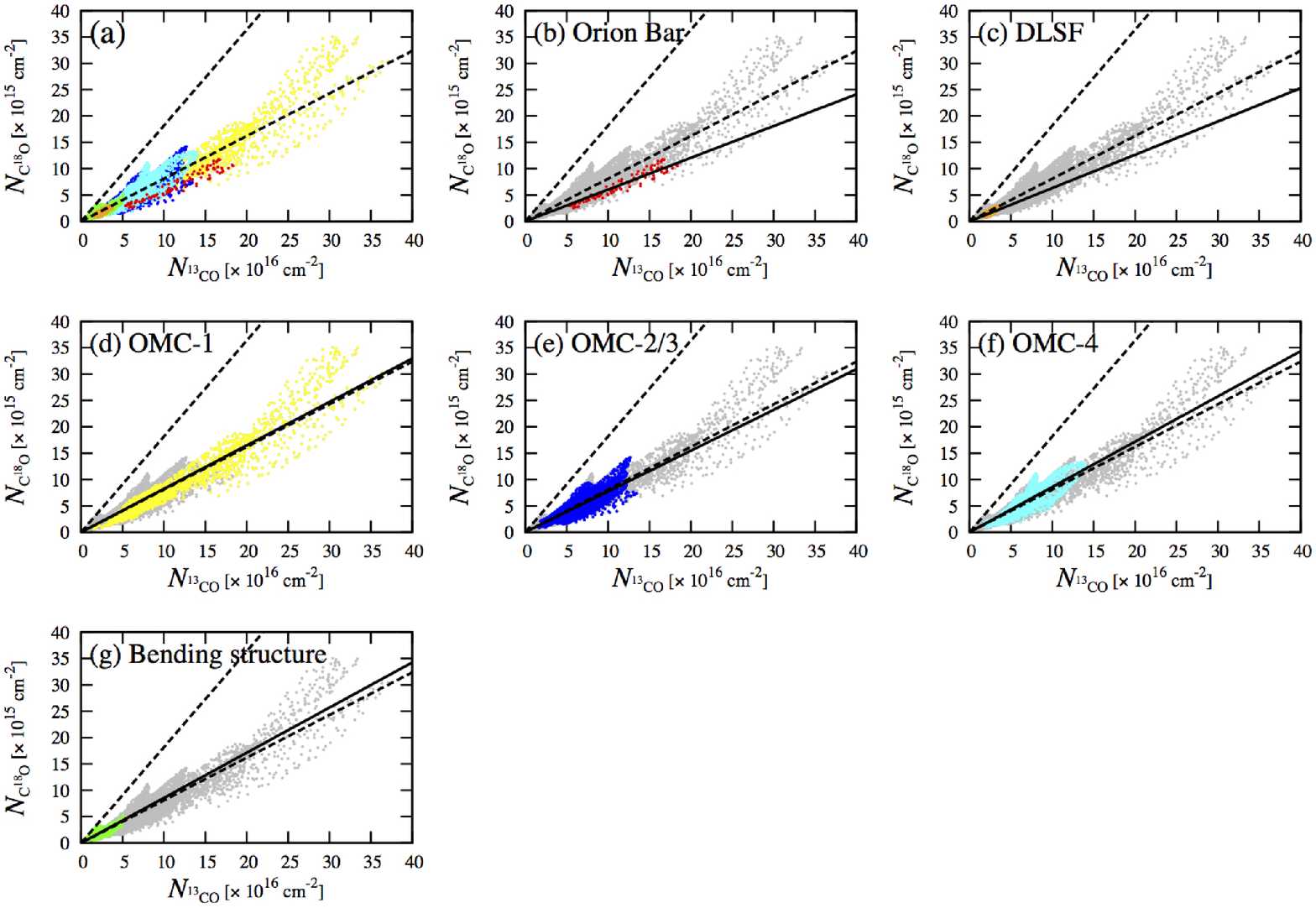}
\caption{
$N_{\rm ^{13}CO}$ vs. $N_{\rm C^{18}O}$ measured in (a) the entire region, (b) the Orion Bar, (c) DLSF, (d) OMC-1, (e) OMC-2/3, (f) OMC-4, and (g) bending structure area,  respectively.
The gray, red, orange, yellow, blue, aqua, and green plots are taken from the entire region, data points in the Orion bar, DLSF, OMC-1, OMC-2/3, OMC-4, and bending structure, respectively. 
The black dashed lines indicate $X_{\rm ^{13}CO}$/$X_{\rm C^{18}O}$=5.5 and \Xall. 
In panels (b) -- (g), black lines denote the best fitting lines to measure the abundance ratio $X_{\rm ^{13}CO}$/$X_{\rm C^{18}O}$ for each region (see Table \ref{ratio_of_each_area}). 
 }
\label{correlation}
\end{center}
\end{figure*}

\begin{figure*}
\begin{center}
\includegraphics[width=170mm, angle=0]{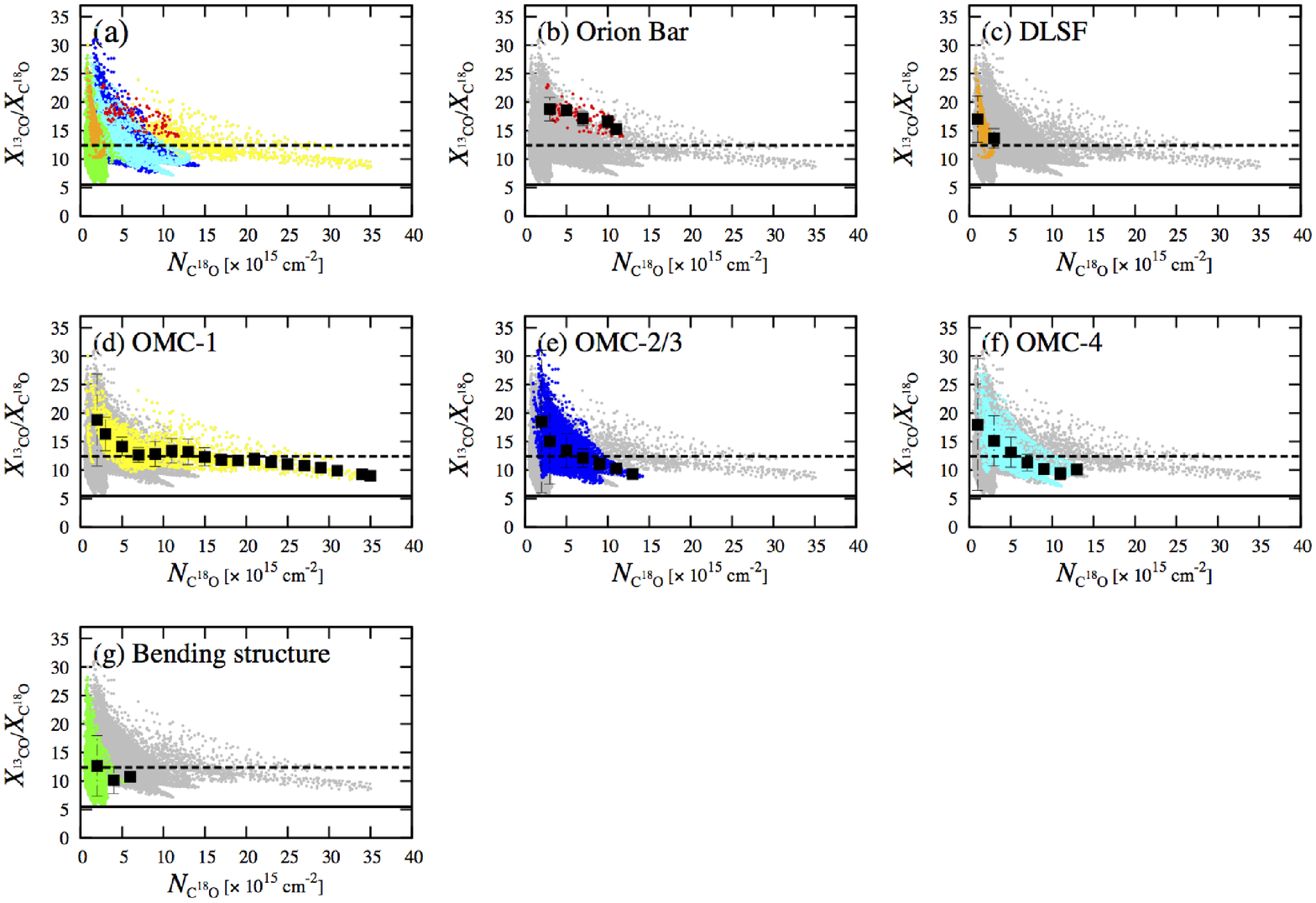}
\caption{
$N_{\rm C^{18}O}$ vs $X_{\rm ^{13}CO}$/$X_{\rm C^{18}O}$ measured in (a) the entire region, (b) the Orion Bar, (c) DLSF, (d) OMC-1, (e) OMC-2/3, (f) OMC-4, and (g) bending structure area,  respectively.
The gray, red, orange, yellow, blue, aqua, and green plots are taken from the entire region, data points in the Orion bar, DLSF, OMC-1, OMC-2/3, OMC-4, and bending structure, respectively. 
The black lines indicate $X_{\rm ^{13}CO}$/$X_{\rm C^{18}O}$=5.5. 
The dashed lines indicate $X_{\rm ^{13}CO}$/$X,_{\rm C^{18}O}$= \Xall, which is the mean value of $X_{\rm ^{13}CO}$/$X_{\rm C^{18}O}$ for the entire region.
The filled squares show the mean $X_{\rm ^{13}CO}$/$X_{\rm C^{18}O}$ which is computed by binning the individually calculated ratios into intervals of 2.0 $\times$ 10$^{15}$ cm$^{-2}$. The error bars show the standard deviations in each bin.}
\label{NXcorrelation}
\end{center}
\end{figure*}

The significant difference between the abundance ratios of $X_{\rm ^{13}CO}$/$X_{\rm C^{18}O}$ in the PDRs and the other regions are clearly demonstrated in the correlation diagrams between the column densities of $^{13}$CO and C$^{18}$O in Fig. \ref{correlation}.
In OMC-1, OMC-2/3, OMC-4, and bending structure, the mean values of $X_{\rm ^{13}CO}$/$X_{\rm C^{18}O}$ are \XOMCone\ $\pm$ \XOMConeerror, \XOMCtwothree\ $\pm$ \XOMCtwothreeerror, \XOMCfour\ $\pm$ \XOMCfourerror, and \XBend\ $\pm$ \XBenderror, respectively. 
On the other hand, in the PDR regions of the Orion bar and DLSF, 
the mean values of $X_{\rm ^{13}CO}$/$X_{\rm C^{18}O}$ are \XBar\ $\pm$ \XBenderror\ and \XDLSF\ $\pm$  \XDLSFerror, respectively.
The $X_{\rm ^{13}CO}$/$X_{\rm C^{18}O}$ values of $\sim$16 is a factor of three larger than the value of 5.5 in the solar system \citep{Wilson92}. 
It should represent the chemical difference between the PDRs and the other regions. 
The chemical difference between the PDR and the other regions is likely to be caused by the different FUV intensities around the regions.
Since the FUV intensities around the PDR regions are considerably higher than those around the other regions, the selective UV photodissociation of C$^{18}$O \citep{Yurimoto04, Lada94} efficiently occurs over the entire PDR region.
Thus, the $X_{\rm ^{13}CO}$/$X_{\rm C^{18}O}$ value should be higher in the PDRs.

The interstellar UV radiation probably also causes the selective UV photodissociation of C$^{18}$O in the outskirts of the cloud. Figures \ref{NXcorrelation} (d), (e), (f), and (g) show that the $X_{\rm ^{13}CO}$/$X_{\rm C^{18}O}$ ratios at low column densities are as high as the ratios in the PDRs, while the ratios decrease with increasing column densities. 
The theoretical study predicts that the $X_{\rm ^{13}CO}$/$X_{\rm C^{18}O}$ ratio reaches values between f5 and 10 in regions with $A_V$ = 1 -- 3 \citep{Warin96}.  
The $X_{\rm ^{13}CO}$/$X_{\rm C^{18}O}$ values at low column densities in the OMC-1, OMC-2/3, OMC-4, and bending structure regions are, however, higher than the predicted values of 5 to 10. 
The reason might be that these regions are influenced by the FUV radiation from OB stars embedded in the Orion-A GMC such as the NGC 1977, M 43, and Trapezium cluster ($\theta^1$ Ori C), as well as the interstellar UV radiation.  
In OMC-1, which is the closest region to the Trapezium cluster ($\theta^1$ Ori C), the $X_{\rm ^{13}CO}$/$X_{\rm C^{18}O}$ ratios are higher than those at the same $N_{\rm C^{18}O}$ column density in OMC-2/3, OMC-4, and bending structure, 
although the ratios decrease with increasing column densities. 

In addition, the $X_{\rm ^{13}CO}$/$X_{\rm C^{18}O}$ ratio remains high even at larger $A_V$ values than the theoretically predicted values of 1 to 3 mag \citep{Warin96}. We can estimate the $A_V$ value from the column density $N_{\rm C^{18}O}$ by using the following relation derived in the Taurus region by \citet{Frerking82}: 

\begin{equation}
{\rm A}_v = \frac{N_{{\rm C^{18}O}} \ ({\rm cm}^{-2})}{2.4 \times 10^{14}} + 2.9.
\end{equation}

\noindent Thus, the value of $N_{\rm C^{18}O}$ = 4.0 $\times$ 10$^{14}$ cm$^{-2}$, which is the lowest value of the C$^{18}$O column density in the observed area, corresponds to the value  of $A_V$ = 4.5 mag. We note that this value of $A_V$ should be the lower limit, because the C$^{18}$O molecules are likely to be selectively dissociated by the FUV radiation. 
We conclude that the FUV radiation penetrates the innermost part of the cloud and the whole of our observed region is chemically influenced by the FUV radiation from the OB stars embedded in the Orion-A GMC. 

The chemical fractionation of $^{13}$C$^{+}$ + $^{12}$CO $\rightarrow$ $^{13}$CO + $^{12}$C$^{+}$ ($\Delta$ E = 35 K) is considered to occur at the cloud surface \citep{Langer84}. The selective photodissociation of C$^{18}$O is, however, thought to be more dominant than the chemical fractionation, because the temperature of the low column density areas ($N_{\rm C^{18}O}$ $<$ 5 $\times$ 10$^{15}$ cm$^{-2}$) of the Orion-A GMC is high ($T_{\rm ex}$ = \Taverage\ $\pm$ \Tstddev\  K).

\subsection{Influence of the uncertainties in the beam filling factor and the excitation temperature on our derived abundance ratio of $^{13}$CO to C$^{18}$O}

\subsubsection{Influence of the beam filling factor \label{filling_section}}

In Sect. \ref{depth}, we derived the optical depths and column densities of $^{13}$CO and C$^{18}$O by using Eqs (\ref{tau13co}) - (\ref{Nc18o}) assuming that the emitting region fills the beam, i.e., the beam filling factors, $\phi_{\rm ^{13}CO}$ and $\phi_{\rm C^{18}O}$, are 1.0. Here, we investigate the influence of the beam filling factor on the derived physical properties. 

According to \citet{Nummelin98} and \citet{Kim06}, we can estimate the beam filling factor by the equation 

\begin{equation}\label{filling}
\centering
\phi =\frac{\theta_{\rm source}^2}{\theta_{\rm source}^2 + \theta_{\rm beam}^2},
\end{equation}

\noindent where $\theta_{\rm source}$ and $\theta_{\rm beam}$ are the source and beam sizes, respectively. 
The effective beam size of the $^{13}$CO and C$^{18}$O data is 25.8$\arcsec$, which corresponds to 0.05 pc at the distance of the Orion-A GMC. 
The sizes of the structures traced by the $^{13}$CO and C$^{18}$O emission lines in the Orion-A GMC can be estimated from previous observations.
The C$^{18}$O emission is thought to trace the dense cores, clumps, and/or filaments. 
On the other hand, the $^{13}$CO emission is likely to trace more extended components than the C$^{18}$O emission as seen in the integrated intensity maps (Figures \ref{mom0_maps} (a) and (b)).
Previous observations toward the Orion-A GMC in H$^{13}$CO$^+$ (1--0) using the Nobeyama 45m telescope identified 236 dense cores with the clumpfind algorithm \citep{Ikeda07}. They estimated the size corrected for the antenna-beam size on the assumption of a Gaussian intensity profile and revealed that the typical size of dense cores traced in H$^{13}$CO$^+$ is 0.14 pc.
The sizes of the dense cores traced in C$^{18}$O are expected to exceed 0.14 pc, since the critical density of H$^{13}$CO$^+$ is higher than that of C$^{18}$O.
As a reference, previous observations toward the Taurus cloud, which is one of the nearest low-mass star forming regions ($d$ = 140 pc, \citet{Elias78}), revealed that the typical size of the dense cores traced in C$^{18}$O is 0.1 pc \citep{Onishi96}. 
Furthermore, the Herschel observations in the 70 $\mu$m, 160 $\mu$m, 250 $\mu$m, 350 $\mu$m, and 500 $\mu$m emissions revealed the omnipresence of parsec-scale filaments in molecular clouds  
and found that the filaments have very narrow widths: a typical FWHM value of 0.1 pc \citep{Arzoumanian11, Palmeirim13}. From the above, the sizes of  the dense cores, clumps, and/or filaments traced in $^{13}$CO and C$^{18}$O are expected to exceed 0.1 pc. 
In the case of our observations, the beam filling factor is expected to exceed 0.8 on the assumption that $\theta_{\rm source}$ $>$ 0.1 pc and $\theta_{\rm beam}$= 0.05 pc. In Tables \ref{Physical_properties} and \ref{ratio_of_each_area}, we summarize the optical depths, the column densities, and the abundance ratios of $^{13}$CO to C$^{18}$O. 
The $X_{\rm ^{13}CO}/X_{\rm C^{18}O}$ values on the assumption of $\phi_{\rm ^{13}CO}$ = 1.0 and $\phi_{\rm C^{18}O}$ = 0.8 decrease compared with those of $\phi_{\rm ^{13}CO}$ = $\phi_{\rm C^{18}O}$ = 1.0.  
Nevertheless, the $X_{\rm ^{13}CO}/X_{\rm C^{18}O}$ values in the nearly edge-on PDRs of Orion Bar and DLSF are (2.3-2.4) times larger than the solar system value.
Furthermore, the $X_{\rm ^{13}CO}/X_{\rm C^{18}O}$ values even in the OMC-1, OMC-2/3, OMC-4, and bending structure are also (1.7-1.9) times larger than the solar system value. 
Even after taking into consideration of uncertainties in the beam filling factors of $^{13}$CO and C$^{18}$O, 
we safely conclude that the abundance ratio of $^{13}$CO to C$^{18}$O is significantly high toward the nearly edge-on PDRs in the Orion-A GMC.

\subsubsection{Influence of the excitation temperature}

To estimate the optical depths and the column densities, we considered the $^{12}$CO ($J$=1--0) peak temperature in $T_{\rm MB}$ as the excitation temperatures of $^{13}$CO ($J$=1--0) and C$^{18}$O ($J$=1--0). 
There is, however, a possibility that the $^{13}$CO and C$^{18}$O lines trace colder areas than the $^{12}$CO line, because the $^{13}$CO and C$^{18}$O lines are optically thinner than the $^{12}$CO line and probably trace the inner parts of the cloud.
\citet{Castets90} found that the $^{13}$CO and C$^{18}$O line emissions trace the regions with a temperature of 20 -- 30 K and $<$ 20 K, respectively,  from the observations toward the Orion-A GMC in the $^{13}$CO (2--1, 1--0) and C$^{18}$O (2--1, 1--0) line emissions with low angular resolutions of 100$\arcsec$ -- 140$\arcsec$. 
Therefore, in order to investigate the influence of the uncertainties in excitation temperature on the derived physical properties, we also derived the physical properties on the assumption of $T_{\rm ex\ ^{13}CO}$ = 30 K and $T_{\rm ex\ C^{18}O}$ = 20 K. 
These values are summarized in Tables \ref{Physical_properties} and \ref{ratio_of_each_area}.

The optical depths obtained with the assumption of $T_{\rm ex\ ^{13}CO}$ = 30 K and $T_{\rm ex\ C^{18}O}$ = 20 K are similar to those for $T_{\rm ex}$ = $T_{\rm ^{12}CO\ peak}$.
In contrast, the column densities derived on the assumption of $T_{\rm ex}$ = $T_{\rm ^{12}CO\ peak}$ are overestimated by a factor of 1.5-3.0. 
The $X_{\rm ^{13}CO}/X_{\rm C^{18}O}$ values are, however, estimated to be a factor of 1.2 -- 1.9 larger than the values obtained with the assumption of $T_{\rm ex}$ = $T_{\rm ^{12}CO\ peak}$.
Thus, the $X_{\rm ^{13}CO}/X_{\rm C^{18}O}$ values in the Orion-A GMC are (2.5 -- 5.0) times larger than the solar system value, even after taking into consideration the uncertainties in the $^{13}$CO and C$^{18}$O  ($J$=1--0) excitation temperatures.

\subsubsection{Robustness of the high abundance ratio of $^{13}$CO to C$^{18}$O}

\begin{figure}
\begin{center}
\includegraphics[width=90mm, angle=0]{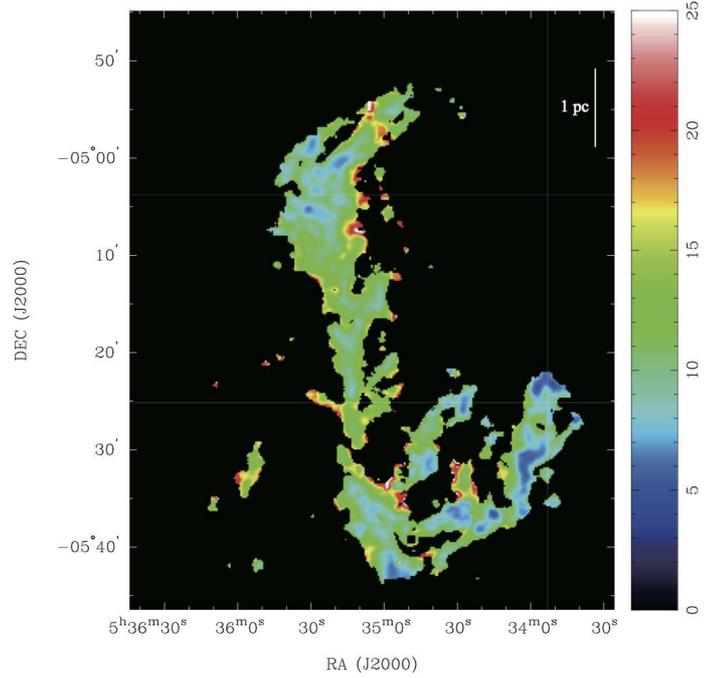}
\caption{Map of the integrated intensity ratio of $^{13}$CO to C$^{18}$O}
\label{integrated_intensity_ratio_map}
\end{center}
\end{figure}

Finally, we demonstrate that the high abundance ratio of $^{13}$CO to C$^{18}$O is most likely to be a direct consequence of the high intensity ratio of $^{13}$CO to C$^{18}$O in the Orion-A GMC.
Figure \ref{integrated_intensity_ratio_map} shows the distribution of the integrated intensity ratio of $^{13}$CO ($J$=1–0) to C$^{18}$O ($J$=1–0), $R_{13,18}$ = $I_{\rm ^{13}CO}$/$I_{\rm C^{18}O}$. The mean $R_{13,18}$ value within the observed area is found to be 11.4 $\pm$ 3.2 (min: 4.9, max:32.3). The $R_{13,18}$ values in the six distinct regions are summarized in Table \ref{integrated_intensity_ratio}. The $R_{13,18}$ values in the nearly edge-on PDRs of the Orion Bar and DLSF seem larger than those in the other regions.
The observed intensity ratio, $R_{13,18}$, is related with the six parameters of $\phi_{\rm ^{13}CO}$, $\phi_{\rm C^{18}O}$, $T_{\rm ex\ ^{13}CO}$, $T_{\rm ex\ C^{18}O}$, $\tau_{\rm ^{13}CO}$, and $X_{\rm ^{13}CO}/X_{\rm C^{18}O}$ (=$A_{13,18}$) as follows:

\begin{equation}\label{intensity_ratio}
\centering
R_{13,18} =\frac{\phi_{\rm ^{13}CO}}{\phi_{{\rm C^{18}O}}} \frac{1-{\rm exp}(-\tau_{\rm ^{13}CO})}{1-{\rm exp}(-\tau_{\rm ^{13}CO}/A_{13,18})} \frac{5.29[J(T_{\rm ex\ ^{13}CO})-0.164]}{5.27[J(T_{\rm ex\ C^{18}O})-0.1666]}.
\end{equation}

\begin{figure*}
\begin{center}
\includegraphics[width=65mm, angle=0]{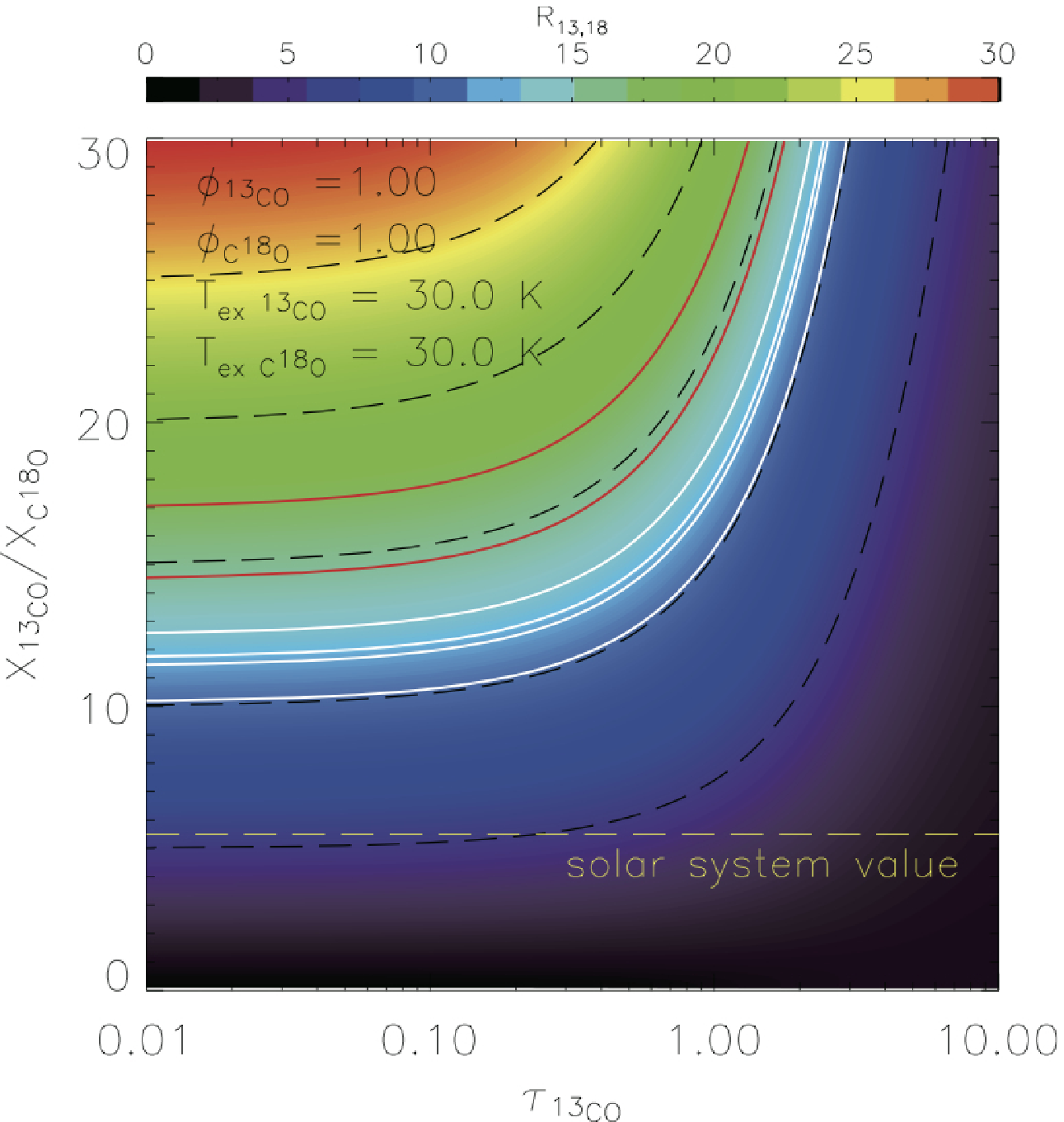}
\includegraphics[width=65mm, angle=0]{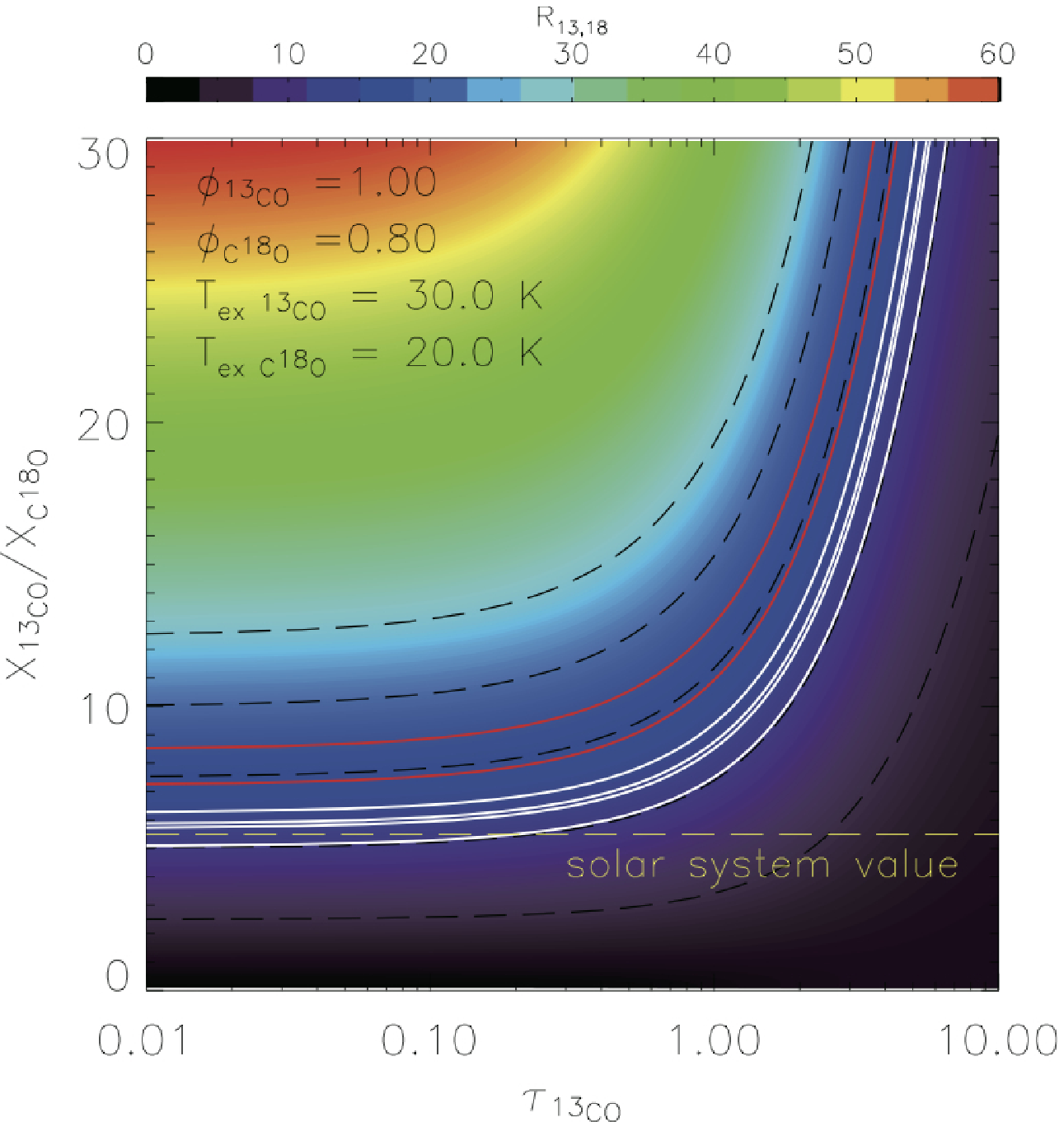}
\caption{Contour maps of the integrated intensity ratio of $^{13}$CO to C$^{18}$O in the parameter plane of the optical depth of $^{13}$CO and the abundance ratio of $^{13}$CO to C$^{18}$O.
In the left panel,  
$\phi_{\rm ^{13}CO}$ = 1.0, $\phi_{\rm C^{18}O}$ = 1.0, $T_{\rm ex\ ^{13}CO}$ = 30.0 K, and $T_{\rm ex\ C^{18}CO}$ = 30.0 K  are assumed.
In the right panel,  
$\phi_{\rm ^{13}CO}$ = 1.0, $\phi_{\rm C^{18}O}$ = 0.8, $T_{\rm ex\ ^{13}CO}$ = 30.0 K, and $T_{\rm ex\ C^{18}CO}$ = 20.0 K  are assumed.
The yellow dashed lines indicate $X_{\rm ^{13}CO}$/$X_{\rm C^{18}O}$ = 5.5, which is the solar system value. 
The black dashed lines indicate $R_{13,18}$ = 5, 10, 15, 20, and 25.
The red lines correspond to $R_{13,18}$ = 17.0 and 14.5, which are the values for the Orion Bar and DLSF, respectively.
The white lines correspond to $R_{13,18}$ = 12.15,11.7,11.4, and 10.1, which are the values for the OMC-1, OMC-2/3, OMC-4, and the bending structure, respectively.
}
\label{relation_filling}
\end{center}
\end{figure*}

Figure \ref{relation_filling} shows contour maps of $R_{13,18}$ in the $\tau_{\rm ^{13}CO}$ - $X_{\rm ^{13}CO}$/$X_{\rm C^{18}O}$ plane for the minimum and maximum values of the coefficient of $\phi_{\rm ^{13}CO)}$/$\phi_{\rm C^{18}O}$ $\times$ $J(T_{\rm ex\ ^{13}CO})$/$J(T_{\rm ex\ C^{18}O})$ in equation (\ref{intensity_ratio}). 
The $R_{13,18}$ value increases with the increasing abundance ratio, if $\tau_{\rm ^{13}CO}$ $<$ $\sim$ 1.
To explain the mean $R_{13,18}$ value of 11.4 within the observed area, the lower limits of the abundance ratio, $X_{\rm ^{13}CO}/X_{\rm C^{18}O}$, are 11.5 for the maximum case in the left panel and 5.8 for the minimum case in the right panel. 
Furthermore, the lower limits of the $X_{\rm ^{13}CO}/X_{\rm C^{18}O}$ values in the nearly edge-on PDRs of the Orion Bar and DLSF are 17.1 and 14.6 for the maximum case and 14.6 and 7.3 for the minimum case. 
In addition, the lower limits of the $X_{\rm ^{13}CO}/X_{\rm C^{18}O}$ values in the OMC-1, OMC-2/3, OMC-4, and bending structure are 12.6, 11.8, 11.5, and 10.1 for the maximum case and 6.3, 5.9, 5.8, and 5.4 for the minimum case. 
In spite of the uncertainties in our adopted parameters in equation (\ref{intensity_ratio}), we conclude that the the lower limits of $X_{\rm ^{13}CO}/X_{\rm C^{18}O}$ values in the nearly edge-on PDRs tend to be larger than those in the other regions and the $X_{\rm ^{13}CO}/X_{\rm C^{18}O}$ values within the observed area are (1.3 - 3.1) times larger than the solar system value of 5.5. 
.5.


\begin{table}
\centering
\caption{Integrated intensity ratio of $^{13}$CO to C$^{18}$O \label{integrated_intensity_ratio}}
\begin{tabular}{lcccc}
\hline
\hline
Region  & $R_{13,18}$ \\
\hline
Orion-Bar & 17.0 $\pm$ 1.9  \\
DLSF &    14.5 $\pm$ 2.7 \\ 
OMC-1 &  12.5  $\pm$ 2.5 \\
OMC-2/3 & 11.7  $\pm$ 3.1 \\
OMC-4  & 11.4  $\pm$ 3.4 \\
Bending structure & 10.1 $\pm$ 3.0\\
\hline
entire region & 11.4 $\pm$ 3.2 \\
\hline
\end{tabular}
\end{table}

\section{Conclusions}

We have carried out wide-field (0.4 deg$^2$) observations with an angular resolution of 25.8$\arcsec$ ($\sim$ 0.05 pc) in the $^{13}$CO ($J$=1--0) and C$^{18}$O ($J$=1--0) emission lines toward the Orion-A GMC. The main results are summarized as follows:

\begin{enumerate}

\item The overall distributions and velocity structures of the $^{13}$CO and C$^{18}$O emission lines are similar to those of the $^{12}$CO ($J$=1--0) emission line. The $^{13}$CO velocity channel maps show a new shell-like structure, which is not obvious in the $^{12}$CO and C$^{18}$O maps. 

\item We estimated the optical depths and column densities of the $^{13}$CO and C$^{18}$O emission lines. The optical depths of $^{13}$CO  and C$^{18}$O are estimated to be  \tauAmin\ $<$ $\tau_{\rm ^{13}CO}$ $<$ \tauAmax\ and \tauBmin\ $<$ $\tau_{\rm C^{18}O}$ $<$ \tauBmax, respectively. 
The column densities of the $^{13}$CO and C$^{18}$O gas are estimated to be \NAmin\ $\times$ 10$^{\NAorder}$ $<$ $N_{\rm ^{13}CO}$ $<$ 3.7 $\times$ 10$^{17}$ and  \NBmin\ $\times$ 10$^{\NBorder}$ $<$ $N_{\rm C^{18}O}$ $<$ 3.5 $\times$ 10$^{16}$ cm$^{-2}$, respectively.

\item The abundance ratio between $^{13}$CO and C$^{18}$O, $X_{\rm ^{13}CO}$/$X_{\rm C^{18}O}$, is found to be \Xmin\ -- \Xmax. The mean value of $X_{\rm ^{13}CO}$/$X_{\rm C^{18}O}$ of the the nearly edge-on PDRs such as the Orion Bar and DLSF are  16.5, which is a factor of three larger than the solar system value of 5.5. 
On the other hand, the mean value of $X_{\rm ^{13}CO}$/$X_{\rm C^{18}O}$ in the other regions is 12.3. 
The difference between the abundance ratios in nearly edge-on PDRs and the other regions are likely due to the different intensities of the FUV radiation that cause the selective photodissociation of C$^{18}$O.

\item In the low column density regions ($N_{\rm C^{18}O}$ $<$ 5 $\times$ 10$^{15}$ cm$^{-2}$), we found that the abundance ratio exceeds 10.  These regions are thought to be influenced by the FUV radiation from the OB stars embedded in the Orion-A GMC such as the NGC 1977, M 43, and Trapezium cluster as well as the interstellar UV radiation.  

\item To examine the influence of the beam filling factor in our observations on the abundance ratio of $^{13}$CO to C$^{18}$O, we estimated the beam filling factors for the $^{13}$CO and C$^{18}$O gas to exceed 0.8. After taking into consideration the uncertainties in the beam filling factor, we also found the high abundance ratio $X_{\rm ^{13}CO}$/$X_{\rm C^{18}O}$ over the Orion-A cloud, particularly toward the nearly edge-on PDRs.

\item 
Even if we consider the lower excitation temperatures of $T_{\rm ex\ ^{13}CO}$ = 30 K and $T_{\rm ex\ C^{18}O}$ = 20 K, we come to the same conclusion that the abundance ratio $X_{\rm ^{13}CO}$/$X_{\rm C^{18}O}$ becomes high toward the nearly edge-on PDRs.

 \item 
 We checked the robustness of our conclusions in the Orion-A GMC by varying $\phi_{\rm ^{13}CO}$, $\phi_{\rm C^{18}O}$, $T_{\rm ex\ ^{13}CO}$, $T_{\rm ex\ C^{18}O}$, $\tau_{\rm ^{13}CO}$, and $X_{\rm ^{13}CO}$/$X_{\rm C^{18}O}$.
 To explain the mean value of 11.4 for the intensity ratio $R_{13,18}$ in our observed region,  the lower limit of the $X_{\rm ^{13}CO}$/$X_{\rm C^{18}O}$ value should be (5.8 -- 11.5) times larger than the solar system value of 5.5. 
In addition, the $X_{\rm ^{13}CO}$/$X_{\rm C^{18}O}$ values in the nearly edge-on PDRs are most likely larger than those in the other regions.

\item
When studying the range of possible values of the beam filling factors and of the excitation temperatures, the conclusion remains valid that the $X_{\rm ^{13}CO}$/$X_{\rm C^{18}O}$ values are higher than solar throughout Orion A, and larger in the PDRs than in the diffuse medium.

\end{enumerate}

\begin{acknowledgements}
We acknowledge the anonymous referee for providing helpful suggestions to improve this paper. The 45-m radio telescope is operated by Nobeyama Radio Observatory, a branch of National Astronomical Observatory of Japan.
This work was supported by JSPS KAKENHI Grant Number 90610551.
Part of this work was supported by the ANR-11-BS56-010 project ``STARFICH". 
\end{acknowledgements}


\bibliographystyle{aa}



\end{document}